\documentclass[12pt,a4paper,final]{iopart}

\usepackage{iopams}  
\usepackage{graphicx}
\usepackage[english]{babel}
\usepackage[utf8x]{inputenc}
\usepackage[T1]{fontenc}

\usepackage{aas_macros}
\usepackage[a4paper,top=3cm,bottom=2cm,left=3cm,right=3cm,marginparwidth=1.75cm]{geometry}

\expandafter\let\csname equation*\endcsname\relax
\expandafter\let\csname endequation*\endcsname\relax
\usepackage{amsmath}
\usepackage{amssymb}
\usepackage{graphicx}
\usepackage[colorinlistoftodos]{todonotes}
\usepackage[colorlinks=true, allcolors=blue]{hyperref}

\newcommand{\be}{\begin{equation}}                                 
\newcommand{\ee}{\end{equation}}                                   
\newcommand{\bea}{\begin{eqnarray}}                                
\newcommand{\eea}{\end{eqnarray}}                                  
\newcommand{\nn}{\nonumber}        
\newcommand{\pp}{\varphi}    
\newcommand{\dd}{\mathrm{d}}
\newcommand{\ui}{\mathrm{i}}

\DeclareMathOperator{\sinc}{sinc}

\begin{document}

\title[Finding IMBH in the black hole desert]{SAGE: finding IMBH in the black hole desert}
\author{S Lacour$^1$, F H Vincent$^1$, M Nowak$^1$, A Le Tiec$^2$, \\
V Lapeyrere$^1$, L David$^1$, P Bourget$^3$, A Kellerer$^3$, \\
K Jani$^4$, J Martino$^5$, J-Y Vinet$^6$, O Godet$^7$,\\
O Straub$^1$ and J Woillez$^3$}

\address{$^1$ LESIA, Observatoire de Paris, Université PSL, CNRS, Sorbonne Université, Univ. Paris Diderot, 5 place Jules Janssen, 92195 Meudon, France}
\address{$^2$ LUTH, Observatoire de Paris, PSL Research University, CNRS, Université Paris
Diderot, Sorbonne Paris Cité, 92190 Meudon, France}
\address{$^3$ ESO, Karl-Schwarzschild-Str. 2, 85748
Garching, Germany}
\address{$^4$ Center for Relativistic Astrophysics and School of Physics, Georgia Institute of Technology, Atlanta, Georgia 30332, USA}
\address{$^5$ APC, Univ Paris Diderot, CNRS/IN2P3, CEA/lrfu, Obs de Paris, Sorbonne Paris Cit\'e, France}
\address{$^6$ Artemis, Université Côte d'Azur, Observatoire Côte d'Azur, CNRS, CS 34229, F-06304 Nice Cedex 4, France}
\address{$^7$ IRAP, CNRS, 9 avenue du Colonel Roche, F-31028 Toulouse Cedex 4, France}

\ead{sylvestre.lacour@obspm.fr}

\date{\today}% It is always \today, today,

%GE paper
\begin{abstract}
SAGE (SagnAc interferometer for Gravitational wavE) is a  project for a space observatory based on multiple 12-U CubeSats in geosynchronous equatorial orbit. The objective is a fast track mission which would fill the observational gap between LISA and ground based observatories. With albeit a lower sensitivity, it would allow early investigation of the nature and
event rate of intermediate-mass black hole (IMBH) mergers, constraining our understanding of the universe formation by probing the building up of IMBH up to supermassive black holes (SMBH). Technically, the CubeSats would create a triangular Sagnac interferometer with 140.000\,km roundtrip arm length, optimised to be sensitive to gravitational waves at frequencies between 10\,mHz and 2\,Hz. The nature of the Sagnac measurement makes it almost insensitive to position error, a feature enabling the use of spacecrafts in ballistic trajectories instead of perfect free fall. The light source and recombination units of the interferometer are based on compact fibered technologies without bulk optics.  A peak sensitivity of 23 pm/$\sqrt{\mathrm{\rm Hz}}$ is expected at 1\,Hz assuming a 200\,mW internal laser source and 10-centimeter diameter apertures. Because of the absence of a test mass, the main limitation would come from the non-gravitational forces applied on the spacecrafts. However, conditionally upon our ability to partially post-process the effect of solar wind and solar pressure, SAGE would allow detection of gravitational waves with strains as low as a few $10^{-19}$ within the 0.1 to 1\,Hz range. Averaged over the entire sky, and including the antenna gain of the Sagnac interferometer, the SAGE observatory would sense equal mass black hole mergers in the $10^4$ to $10^6$ solar masses range up to a luminosity distance of 800\,Mpc. Additionally, coalescence of stellar black holes (10\,M$_\odot$) around SMBH (IMBH) forming extreme (intermediate) mass ratio inspirals could be detected within a sphere of radius 200\,Mpc.
\end{abstract}

\submitto{\CQG}


\maketitle

\section{Introduction}

\begin{figure}[h]
	\centering
	\includegraphics[width=0.75\textwidth]{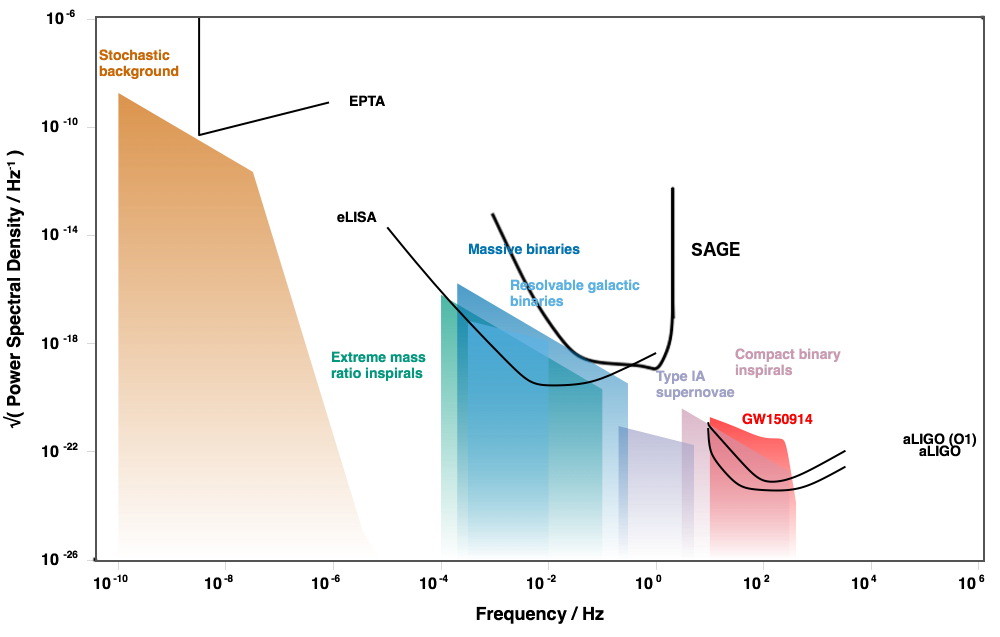} 
	\caption{Sensitivity curves of gravitational-wave observatories~\cite{moore15}.}
	\label{fig:sensitivity}
\end{figure}

Gravitational-wave astronomy will be a major observing window on the Universe 
in the next decades. Fig.~\ref{fig:sensitivity} shows the sensitivity curves of
the existing and planned gravitational-wave observatories. The ground-based detectors that obtained
the first direct detection of gravitational waves~\cite{abbott16} are sensitive around $10^3$~Hz,
while the future Laser Interferometer Space Antenna (LISA) 
will reach maximum sensitivity at $\approx 10^{-2}$~Hz \cite{2017arXiv170200786A}.
The gap in between these two values is and has been the subject of investigation
by various groups, with quite a few space interferometer proposals published
in the literature (for a review of gravitational wave detection in space, see reference~\cite{ni16}).
These proposals can be divided in helio- and geocentric instruments. 

In the heliocentric category,
the Chinese \textit{ALIA}~\cite{gong15} was initially designed to reach a much better sensitivity (by $\approx 2$ decades) than LISA
in the $0.1 - 1$~Hz range, in order to detect intermediate-mass black holes.
It is made of a spacecraft triangle of side $3\times10^6$~km, with $\approx 0.5$~m diameter
telescopes, and 2~W lasers.
It was descoped in 2015 and is now planned to perform only slightly better than LISA in this range. The American \textit{BBO}~\cite{harry06}
is designed to study the gravitational-wave cosmological background in the early universe.
It should reach an accuracy $\approx 4$ orders of magnitude better than LISA in the
$0.01 - 1$~Hz region.
It is made of a triangle of 3 spacecrafts with arm length $5\times10^4$~km, $2.5$~m mirrors,
and $300$~W lasers. The Japanese \textit{DECIGO}~\cite{kawamura11,doi:10.1142/S0218271818450013} has a sensitivity $\approx 3$
orders of magnitude better than LISA in the deci-Hz region (see Fig~\ref{fig:sensitivity}). 
It is made of 4 clusters of 3 spacecrafts
distributed on triangles with arm length $1000$~km, $1$~m mirrors, and $10$~W lasers.
It has a broad science case, comprising in particular the characterization of inflation,
the formation mechanism of supermassive black holes, and tests of gravitation. 
A first-step instrument, Pre-DECIGO, as described in Ref.~\cite{nakamura16}, is 
announced for the late 2020s.

Downscaled, geostationary LISA-like missions (with $73000$~km arms) have been proposed in response to 
a 2011 NASA call for gravitational-wave detector concepts: \textit{GEOGRAWI}~\cite{tinto11} and \textit{GADFLI}~\cite{mcwilliams11},
two very similar concepts developed independently for the call.
These designs have the obvious advantage of being less expensive, and more sensitive than LISA in the deci-Hz region by up to a factor 10. 
The mirror size can be as small as $15$~cm
with $0.7$~W lasers. The main science objective is the follow-up of massive
black hole mergers, with masses smaller than the LISA band.
Such missions have been collectively named \textit{gLISA}
(geosynchronous LISA) in a follow-up study~\cite{tinto15},
which also proposes to fly the instrument on industrial telecom satellites
to further reduce the costs. 
The \textit{LAGRANGE} proposal~\cite{conklin11} is made of triangles of spacecrafts
located at 3 Lagrange points of the Earth-Moon system, which is the most stable
geocentric orbit. The arm length is of $670 000$~km, with $20$~cm mirrors, and $1$~W lasers.
The sensitivity in the deciHz region is similar to that of the gLISA class.
The main objectives of LAGRANGE are the follow-up of massive black hole
mergers, extreme mass-ratio inspirals, and galactic binaries.
The recent Chinese \textit{TianQin}~\cite{luo16} proposal is based on a different
approach: it is optimized for one particular source, the galactic binary RX J0806.3+1527,
which is the strongest known periodic emitter of gravitational waves in the deci-Hz region.
The instrument is made of a triangle of spacecrafts in Earth orbit with $10^5$~km
arm length, $20$~cm mirrors, and $4$~W lasers. The sensitivity is similar
to that of the gLISA class.

%\red{Among all these missions, only descoped ALIA, DECIGO, gLISA and TianQin seem still alive.
%Probably gLISA is the SAGE-like competitor, submitted to NASA, but I think with no clear funding. 
%DECIGO is of much broader scale, and a tough competitor. Pre-DECIGO, announced for
%the late 2020s, would probably already kill SAGE (to be checked). This would mean launching something
%like mid 2020s.}

This paper presents a new downscaled instrument of the gLISA class: SAGE (SagnAc interferometer for Gravitational wavE) is made of a triplet
of identical CubeSats in geosynchronous equatorial orbit (GEO). Our aim is to show what can be obtained through use of CubeSats, with a limited total power budget of the order of 20\,W, off-the-shelves fiber laser sources, laser thermal dilatation of conventional materials, aperture sizes achievable with CubeSats, limited post-processing capabilities, etc... We find that the detection limits are a magnitude higher than LISA. However, the maximum sensitivity is designed to lie 
around $1$~Hz, in order to bridge the gap between LISA and ground-based
observatories (Figure~\ref{fig:sensitivity}). With an estimated cost below 100\,M\$, and a development time between 5 and 10\,years, the SAGE CubeSats could attempt to detect GW at intermediate frequencies and thus pave the way for more ambitious space programs.

The paper is structured as follows: section~\ref{sec:science} presents the science objectives centred around IMBH.
Section~\ref{sec:detector} introduces the detector design. Section~\ref{sec:error}
reviews the different noise sources, while Section~\ref{sec:sensitivity} computes the sensitivity function of the instrument. Finally, Section~\ref{sec:summary}
concludes on the use of CubeSats and their advantage compared to larger missions.

\section{Science Goal}
\label{sec:science}

\subsection{The observation window}

\begin{figure*}
\centering
\includegraphics[width=0.75\textwidth]{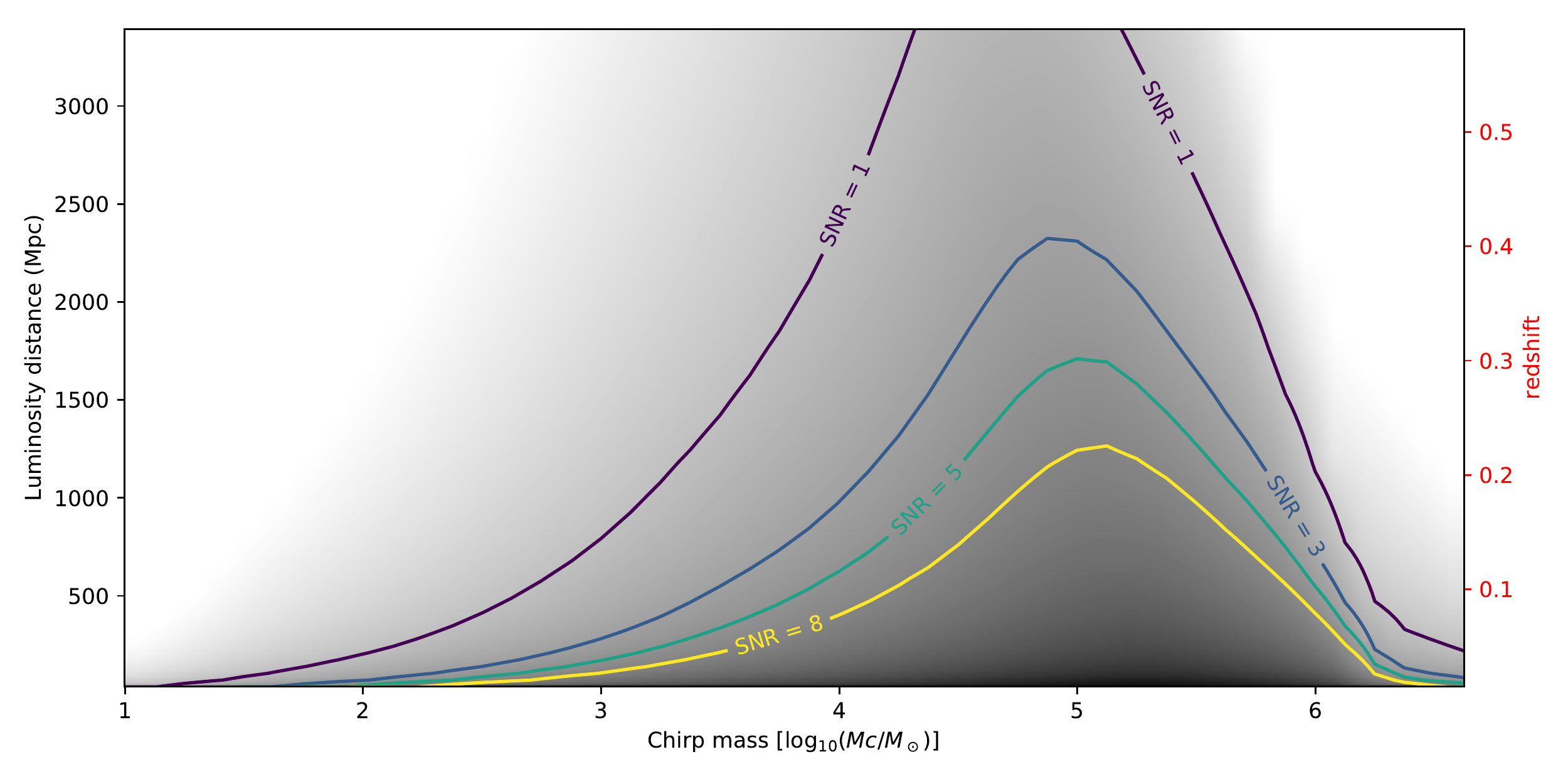}
\caption{Sensitivity as a function of mass and distance for equal mass binary mergers. The computation of the SNR is described in Section~\ref{sec:sensitivity} by Eq.~(\ref{eq:snr2}). It includes the transfer function of the noise and the response of the antenna in the case of an un-equal arm interferometer measurement (8-pulse response). The yellow contour corresponds to an SNR of 8, and reaches 1\,Gpc for a chirp mass of $10^5\,M_\odot$. Over the $10^4$ to $10^6\,M_\odot$ band, the mean horizon luminosity distance is 800\,Mpc. Chirp masses are physical masses, they are not redshifted. The observed chirp mass would be the redshifted mass $\mathcal{M}_c = M_c(1+z)$ \cite{2017ogw..book...43C}. \label{fig:massSensitivity}}
\end{figure*}

The sensitivity of the SAGE interferometer, as established in Sections~\ref{sec:error} and \ref{sec:sensitivity}, writes:
\begin{equation}
S_{\rm h}(f)=\frac{ |H_{\rm noise}(f)|^2}{12|H_{\rm GW (0)}(f) |^2 L^2} \sum_{\rm all\ noises} N(f)\,.
\end{equation}
where $L=73\,000\,$km, $H_{\rm GW (0)}(f)$ is spelled in Eq.~(\ref{eq:Hgw}), $H_{\rm noise}(f)$ in Eq.~(\ref{eq:Hnoise}), and $\sum N(f)$ corresponds to the power spectrum of the noises listed in Table~\ref{tb:noise}. From this sensitivity curve, Fig.~\ref{fig:massSensitivity} shows the sensitivity of SAGE to equal-mass binaries as a function of
mass and luminosity distance. 
SAGE specializes on
a rather narrow black hole mass range, between $10^4$ and $10^6\,M_\odot$,
i.e. in the middle of the black hole "desert" as coined by Ref.~\cite{Auger17}.
These sources are detectable out to redshift $z = 0.3$ (with maximum
redshift obtained for $10^5\,M_\odot)$.
Stellar-mass black holes (like GW150914) and supermassive black holes (above
few times $10^6\,M_\odot$) are only detectable at distances of the order of
tens of Mpc, leading to negligible merger probability. Consequently, this science
case focuses on intermediate-mass black holes (IMBH) in the local Universe, 
a class of compact sources that
has never been detected. 

\subsection{The IMBH population}

Black holes with masses below $100\,M_\odot$ and above $10^6\,M_\odot$
are standard astrophysical sources that can be observed 
in X-ray binaries for the first class, or at the centers of
galaxies for the second class. The existence of IMBH lying in between those
two classes, i.e. with mass $100\,M_\odot < M < 10^6\,M_\odot$, 
is the object of intense debate in the community~\cite{mezcua17,koliopanos17}.
Such IMBH in the local Universe could either have grown in dense
stellar clusters following the
evolution of their host galaxies,
or could be the unevolved remnants of the
initial building blocks of 
supermassive black holes (SMBH) at redshift $z \approx 15 - 20$. 
These SMBH seeds could be either of moderate masses $\approx 100 \,M_\odot$
("light-seed" hypothesis) if they are the leftover of the collapse of massive Pop III stars, or of high masses $\approx 10^5\,M_\odot$ ("heavy-seed" hypothesis) if they are the result of the direct
collapse of metal-free gas halos in protogalaxies.
Demonstrating one of these scenarios (or invalidating both of them)
is a key goal of cosmology.
Although directly observing 
IMBH in the early universe is not within reach of the current
instrumentation, leftover IMBH, that did not grow
to SMBH, might be observable in the local Universe up to a redshift
of around $z\approx 2.4$~\cite{mezcua18}. These sources
could be detected at the center of nearby dwarf galaxies (that have not
significantly evolved since the early Universe and might thus still harbor
their seed IMBH), 
in close gobular clusters, stellar clusters~\cite{lin18}, or as off-nuclear, ultraluminous
X-ray sources. The most likely IMBH candidate as of today is
arguably the hyperluminous X-ray source HLX-1~\cite{farrell09}.

\subsection{IMBH - equal-mass black hole inspiral}

\begin{figure}
\centering
\includegraphics[width=0.75\textwidth]{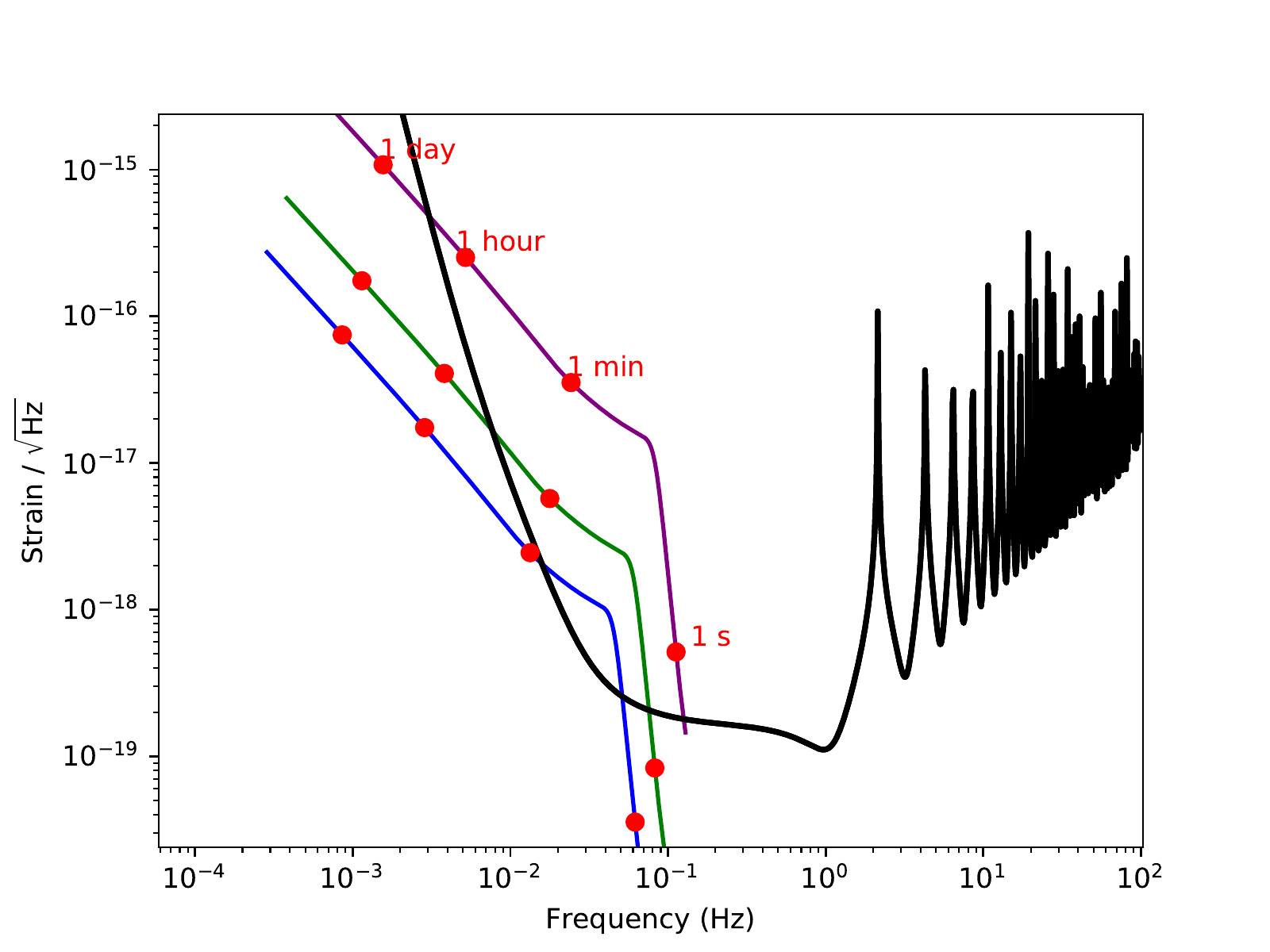}
\caption{Gravitational-wave signal of an equal-mass binary with individual mass $m=10^5\,M_\odot$ at redshift $z=0.1$ (purple), $z=0.5$ (green) or $z=1$ (blue). The corresponding SNR are $15$, $1.5$, $0.4$. A few values of the time to merger are depicted in red. The sensitivity of SAGE is shown in black.
\label{fig:IMBHcoal}}
\end{figure}

Figure~\ref{fig:IMBHcoal} shows the gravitational-wave signal of an equal-mass binary with individual mass $m=10^5\,M_\odot$, as compared to the sensitivity of SAGE. Here, we consider zero-spin black holes in circular orbits, seen under an inclination of $1$~rad. We use the open-source software PyCBC~\cite{Canton:2014ena,Usman:2015kfa,Nitz:2017svb,alex_nitz_2018_1256897} to compute the waveform, considering the template EOBNRv2\_ROM~\cite{pan11}, that uses effective one-body methods tuned to numerical relativity. The corresponding signal-to-noise ratio is above $10$ (resp. $5$) for a redshift of the source below $z\approx 0.14$, corresponding to an age of the Universe of $t = 11.9~Gyr$ (resp. $z\approx 0.23$ or $t = 10.9~$Gyr). 

The merger rate of IMBH depends on their formation and accretion history. The most standard hypothesis is to consider that they are seeds to massive black holes along a hierarchical build up which  result in reproducing the observed active galactic nucleus optical luminosity function (LF) at $z<6$ \cite{2001AJ....122.2833F}. The evolution models have many unconstrained parameters: the seed masses,  the accretion rate, and the coalescence efficiency. Moreover, the constraint caused by the LF underestimate the MBH seed as the source of black hole growth for mergers \cite{2006ApJ...639....7K,sesana07}. In the most optimistic models, according to Ref.~\cite{sesana07} (their Fig.~1), the merger rate for $10^4\,M_\odot < M < 10^6 \, M_\odot$ equal-mass binaries at redshift $z<0.3$ is $< 0.05$ occurrence per year: at best, SAGE would detect an event every 20 years. This is an expected result since the theory assumes that IMBH assembled early in the history of the universe and then progressively merged into more massive black holes. Using the models presented in~\cite{2012MNRAS.423.2533B,klein16,bonetti18}, the averaged merger rate for $10^4\,M_\odot < M < 10^6 \, M_\odot$ equal-mass binaries for
$z<0.3$ lies between $0.04$ and $0.34$ (E. Barausse, private communication).

However, as far as observations are concerned, very little is known about the merger rate of IMBH. By analyzing the data from the first observing run (O1) of the Advanced LIGO detectors, upper limits have been set on the merger rates of comparable-mass black holes in the $100-300 M_\odot$ range. For instance, the merger rate for $100\,M_\odot+ 100\,M_\odot$ (resp. $300\,M_\odot + 300 
\,M_\odot$) nonspinning black holes is constrained to be less than $2.0~\text{Gp}^{-3} \cdot \text{yr}^{-1}$ (resp. $20~\text{Gp}^{-3} \cdot \text{yr}^{-1}$) at the 90\% confidence level \cite{2017PhRvD..96b2001A}. But the low sensitivity of Advanced LIGO and Advanced Virgo below 10~Hz does not allow those detectors to set astrophysically relevant constraints over the $10^4$ to $10^6 M_\odot$ range, for which the mean horizon luminosity distance of SAGE is 800~Mpc (see Fig.~\ref{fig:massSensitivity}). Therefore, provided that the event rate for such IMBH mergers is no less that $2~\text{Gp}^{-3} \cdot \text{yr}^{-1}$, SAGE could observe on average one such event per year or more. Alternatively, SAGE could set an upper bound on the merger rate of IMBH in the unexplored $10^4$ to $10^6 M_\odot$ range.

%It is plausible that electromagnetic radiation could be observed at the same tiem as the binary merger, assuming that the IMBH are surrounded by an accretion flow. This  

\subsection{IMBH - stellar-mass black hole inspiral}

Extreme-mass-ratio-inspirals (EMRIs), consisting of a stellar-mass compact
object spiralling towards a massive companion, are astrophysically rich 
events~\cite{amaro18}. In this Section, we are interested in determining
whether SAGE can observe such radiation, focusing on a $10\,M_\odot$
black hole spiralling towards an IMBH of $10^5\,M_\odot$.

We use a semi-analytic model describing the gravitational-wave emission
from a point source in a Keplerian orbit around a massive primary.
The waveform from this setup is a standard result and reads~\cite{Te.73} %,De.78,Sh.94,Ke.98,Hu.00,GlKe.02}
\be\label{h}
    h = \frac{2\mu}{D} \, \sum_{\ell=2}^{\infty} \sum_{\genfrac{}{}{0pt}{}{\scriptstyle m=-\ell}{ \scriptstyle m\not=0}}^\ell \frac{Z^\infty_{\ell m}(r_0)}{(m\omega_0)^2} \, _{-2}S^{am\omega_0}_{\ell m}(\theta,\varphi) \, e^{ \ui m \omega_0 u}\,\Pi_{T}(u-\frac{T}{2}) ,
\ee
where $\mu = 10\,M_\odot$ is the point-source mass, $D$ is the luminosity distance
to the source, $r_0$ is the Boyer-Lindquist coordinate radius of the circular orbit, 
$\omega_0 = 1/(r_0^{3/2}+a)$ is the associated Keplerian orbital frequency, with $a$ being the massive black hole spin,
$Z^\infty_{\ell m}(r_0)$ encodes the amplitude of each $(\ell,m)$ mode, 
$_{-2}S^{am\omega_0}_{\ell m}(\theta,\varphi)$ are the spin-weighted spheroidal harmonics, and $u$ is the retarded time. 
We have here added a rectangular function, $\Pi_{T}(u-\frac{T}{2})$, defined such that the waveform is nonzero only for retarded times satisfying $0 < u < T$,
where $T$ is the lifetime of the mission.
The Fourier-domain waveform reads
\be
\tilde{h} = \frac{2\mu}{D} \, \sum_{\ell=2}^{\infty} \sum_{\genfrac{}{}{0pt}{}{\scriptstyle m=-\ell}{ \scriptstyle m\not=0}}^\ell \frac{Z^\infty_{\ell m}(r_0)}{(m\omega_0)^2} \, _{-2}S^{am\omega_0}_{\ell m}(\theta,\varphi) \,T\,\mathrm{sinc}\left[(f - \frac{m\omega_0}{2\pi}) T\right]\,e^{-\ui\pi (f-\frac{m\omega_0}{2\pi}) T},
\ee
where the sinus cardinal is defined as $\sinc(x)=\sin{(\pi x)}/\pi x$. We further simplify by setting $\varphi=0$ (the spacetime being axisymmetric, we do not lose generality) and integrating over $\theta$ by replacing
\be
_{-2}S^{am\omega_0}_{\ell m}(\theta,\varphi) \rightarrow \frac{1}{2} \int_0^\pi \! _{-2}S^{am\omega_0}_{\ell m}(\theta,0) \, \sin\theta \, \dd \theta \,.
\ee

This expression allows to readily compute the signal-to-noise ratio,
assuming that the small compact body stays at a constant radius $r_0$.
However, due to radiation reaction, the small body will slowly drift
towards the IMBH, over a characteristic time that is small compared to the
mission lifetime. To obtain a simple estimate of the SNR,
we use a Newtonian treatment to determine the evolution $r_0(u)$ of the orbital radius
with time. 
We find that, considering an IMBH of spin $0$,
mass $M=10^5\,M_\odot$, a small body of mass $\mu=10\,M_\odot$,
and a mission lifetime of 1 year, the small body will drift from
$r_0=24M$ to the innermost stable orbit at $r_0 = 6M$ within
the mission lifetime. These two values of radius allows us to derive upper
and lower bounds for the distance at which such an EMRI would
be detectable. We find that the SNR gets larger than $5$  provided that the EMRI is closer than $15 - 200$~Mpc (redshift $z<0.003 - 0.045$) depending
on the radius evolution. 

Reference~\cite{fragione18} predicts the rate of such EMRI events to be of a few per year per Gpc$^{-3}$ at redshift $0$, so a factor $\approx 100$ less than that in a sphere of radius $200$~Mpc, which is very small. However, the authors mention that this rate might be increased a lot if IMBH are present in young massive stellar clusters.

\section{The detector}
\label{sec:detector}

\subsection{Interferometer geometry}

\begin{figure}
\centering
\includegraphics[width=0.45\textwidth]{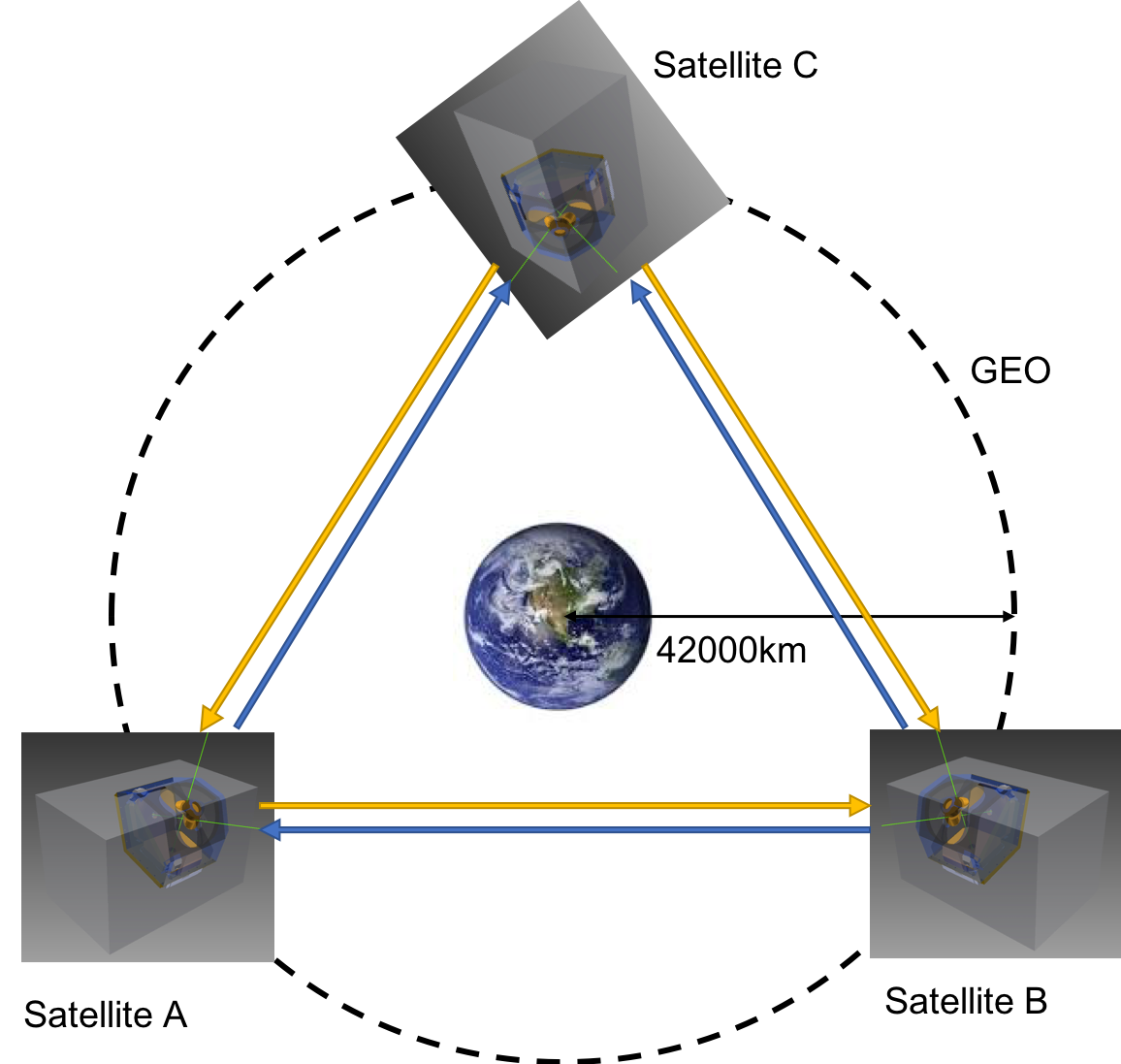}
\caption{ Representation of the  optical setup. All satellites are identical 12-U CubeSats. Each satellite is at an orbit close to GEO (in the so-called graveyard orbit). Together, they form an equilateral triangle of 73\,000\ km arms.
Six laser beams are used to close the triangle, and TDI interferometry is used for the optical path-length measurement.
\label{fig:constel}}
\end{figure}

The goal of the SAGE antenna is to open the window for $\approx 1$\,Hz gravitational waves. 
The main idea is to use three identical 12U CubeSats at geosynchronous orbits. The dimension of the CubeSats ($34\times22\times22\,$cm) forces the simplicity of the mission to its maximum: the satellites are kept on a ballistic trajectory and work without internal gravitation reference sensors. 
This implies that the spacecrafts must, by themselves, act as test masses. The measurement is done between the center-of-mass positions of the different satellites, as it was done for the GRACE and GRACE-FO experiments~\cite{2016SGeo...37..453F}.

This is only possible because solar radiation and wind pressure are small enough at these mid-frequencies~\cite{2018arXiv180608106L}.
However, on top of the gravitational force, the center of mass is affected by pointing errors and by thermal expansion of its material: the satellites have to be steady and compact, and thermally invariant. 
This does prevent the use of deployable solar panels, limiting the energy power budget to a few tens of Watts.
But the concept of SAGE relies on the increasing capabilities of nano-satellites. These platforms are already succesfully used in the geostationnary arc, where they compete with conventional geostationary telecom missions. 
A major difficulty is the deployment into GEO, which requires dedicated propulsion systems. The industry is slowly lifting this barrier, with proposals for bigger satellites to carry small satellites towards near-GEO orbits. 

In the case of SAGE the satellites will go directly in the so-called graveyard orbit \cite{2005ESASP.587..373J} from where no deorbitation would be necessary. 
Each spacecraft is at an altitude of around 36\,000\,km, with the full interferometer forming an equilateral triangle of side 73\,000\,km (Figure~\ref{fig:constel}). 
An internal fine pointing system will account for any drift up to $0.1^\circ$ of the nominal $60^\circ$ angles. This formation  
can be passively maintained for 15 consecutive days \cite{2015CQGra..32r5017T} until small electric thrusters will have to be activated to restore the constellation.
A summary of the mission parameters are listed in Table~\ref{tb:param}. 

\subsection{Telescope design}
\label{sec:telescopes}

\begin{table}
   \caption{\label{tb:param} Mission parameters}
   \centering
   \begin{tabular}{|l|c|}
     \hline
     \textbf{ Satellites}&\\
     \hline
     Mission duration:& $\geq  3$\ years \\
     Number of spacecrafts:& $N=3$\\
     Configuration:& CubeSat (12U)\\
     Size (per satellite):& $34\times22\times22\,$cm \\
     Mass (per satellite):& $M=20\,kg$ \\
     Thermal expansion:& $\alpha=2\times10^{-6}\, K^{-1}$ \\
     Science telemetry:& $120\,$Octets/s\\
     \hline
     \hline
     \textbf{ Geometry}&\\
     \hline
   Constellation:& Equilateral \\
   Orbit:& Ballistic over 15 days \\
   Arm Length:& $L=73\,000$\,km \\
   GW Frequency: & $10$\,mHz $ <f<2$\,Hz\\
   inter-spacecraft velocity:& $v<1\,$m/s \\
   inter-spacecraft angle:& $60^\circ\pm15"$\\
     \hline
     \hline
     \textbf{ Metrology system}&\\
     \hline
   Telescope Diameter:& $D=10$\,cm\\
   Laser type:& Pigtailed External Cavity\\
   Laser wavelength:& $\lambda=1.55\,\mu$m, tunable\\
   Laser power:&  $P_{\rm laser}=200$\,mW\\
   Frequency noise:& $F_{\rm laser} = 20\,$kHz/$\sqrt{\rm Hz}$\\
   Detector bandpass:& 1\,GHz\\
   Measurement method:& Sagnac, TDI \\
   Clock stability (ADEV): & $\sigma_A=10^{-11}$\\ 
     \hline
     \hline
     \textbf{ Requirements}&\\
     \hline
   Pointing accuracy& $0.13^\circ$\\
   Pointing precision& 50\,mas/\,$\sqrt{\rm Hz}$\\
   Solar wind monitoring& 10\% @ 1\,Hz\\
   Solar irradiance monit.& 10\% @ 1\,Hz\\
   Temperature regulation:& $T_{\rm stable}= 0.1\,$K \\
   \hline
   \end{tabular}
\end{table}

Between each observation sequence, the satellites manoeuvre to maintain the equilateral formation. The observation sequences are expected to last 15 days.
In a geostationary orbit, it was demonstrated that the $60^\circ$ formation can be kept during the observation sequence within 5 arc-minutes~\cite{2015CQGra..32r5017T}. This means that the formation can be kept in place with high accuracy, and the satellites do not require  moving optical parts. Thus, for compactness, but most importantly, for robustness, the two telescopes are intertwined. The two primary mirrors, of size 10\,cm, are maintained together by molecular cohesion. The two secondary mirrors, of diameter 36\,mm, are also similarly glued and maintained together by molecular cohesion. To maximize thermal stability, the two M2s are kept in position with respect to the M1s through a common structure of low thermal expansion material.

For thermal stability of the angular reference, on axis optics was preferred at the cost of a central obstruction of 14\% of the total throughput. Both M1 and M2 units are integrated and glued into the invar monobloc body of the telescope. The invariant point of the differential thermal expansion of the complete assembly is as close as possible to the M2 unit. An accurate temperature sensing all over the two telescopes assembly provides input to the model of the differential thermal expansion (Invar/SiC/Zerodur) on the $60^\circ$ reference pointing. 

The only moving part of the optical system are the single mode fibers which emit the beams from the focal planes. This fine pointing corrects for the slow $5'$ drift of the constellation angle. The fiber termination is mounted on a 2 axis piezo stage with a range of  900\,$\mu$m. This range, combined to the f/D value of 0.25, gives a field of view of $0.13^\circ$ ($7.7'$). This means that the constellation angles must be kept within $7.7'$ of the nominal $60^\circ$. %It also gives the requirement on the pointing accuracy of the satellites.

\begin{figure}
\centering
\includegraphics[width=0.8\textwidth]{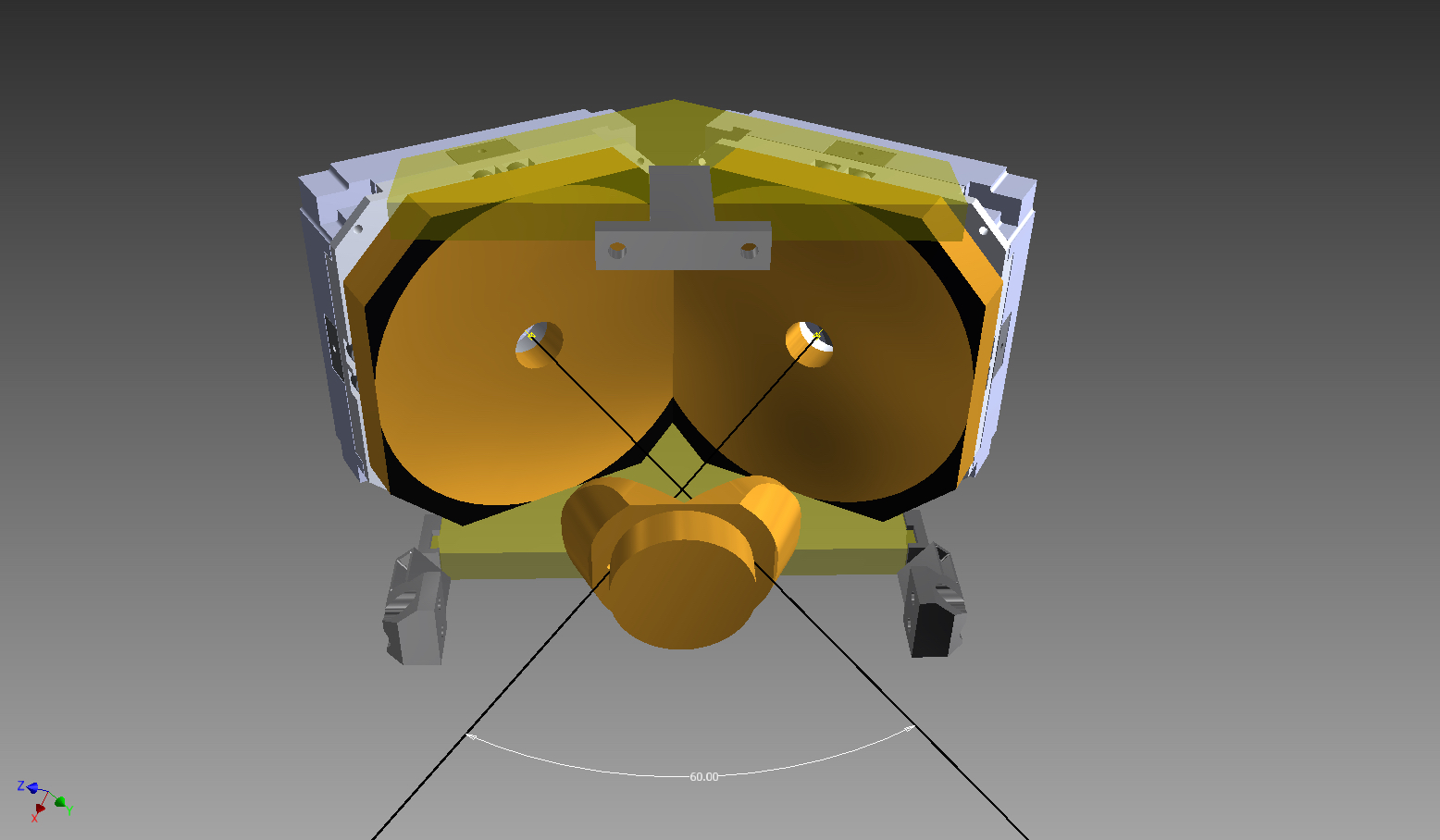}
\caption{3D rendering of the optical part of the spacecraft. The two telescopes are intertwined at $60^\circ$, with the two optical axis represented as the black lines. Fine pointing is obtained thanks to two piezo stages which control the position of the single mode fiber (the piezo stages are the two gray boxes). In direct contact with the piezo stages are the two M1 mirrors, of 10\,cm diameter. On axis of the two M1 mirrors are the two M2 mirrors, centered on the pupil of each telescopes. The two primary and the two secondary mirrors are maintained together by molecular cohesion. 
\label{fig:opto}}
\end{figure}

\begin{figure}
\centering
\includegraphics[width=0.99\textwidth]{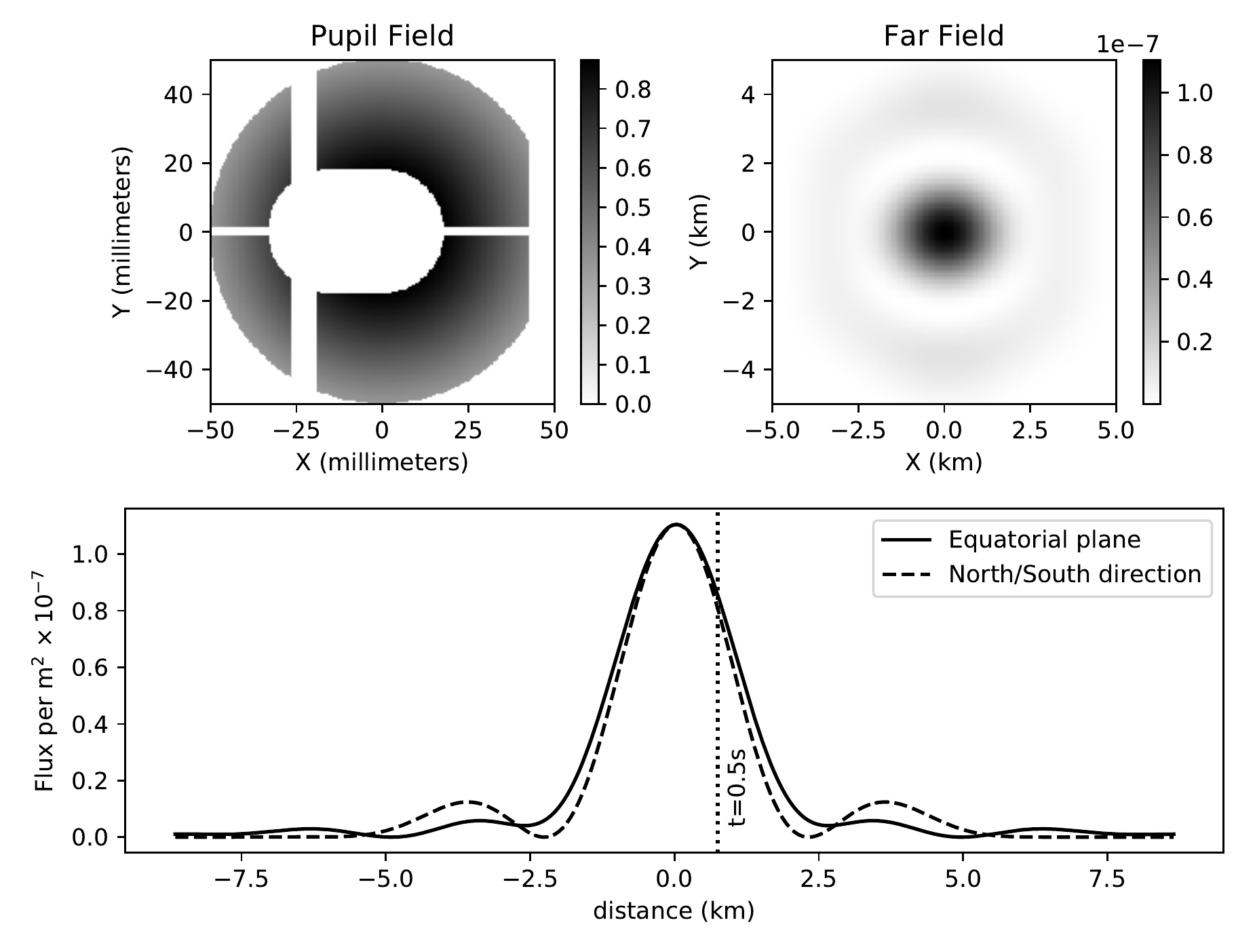}
\caption{ \textit{Upper-left panel:} the pupil including the central obstruction and the shadow caused by the spider arms and the secondary mirror. \textit{Upper-right panel:} the diffraction pattern at a distance of 73\,000\,km, in flux per square meter. Due to the geometry of the pupil, the diffraction pattern is not circular (as an Airy pattern would be). The full width at half maximum of the pattern is of the order of 2\,km. \textit{Lower panel:} X and Y-cut of the diffraction pattern. The 0.5\,s delay between the moment when the light is emitted and when the light is received means the spacecraft will be positioned on the vertical dotted line at 0.75\,km.
\label{fig:diff}}
\end{figure}

The telescopes will not try to point ahead of the position of the other spacecrafts: each fiber will be positioned in the focal plane to  maximize the incoming flux. The fibers will therefore launch a beam toward a satellite as it was 0.25 seconds earlier, and by the time the other satellite will receive the beam it will have moved by 1.5\,km.  Projected toward the other satellite, this distance corresponds to 0.75\,km. This projected distance is plotted as the dotted line in Fig.~\ref{fig:diff}. It can be compared to the width of the diffraction pattern which is above 2.5\,km along the Equatorial direction.

\subsection{Fibered optical bench}

\begin{figure*}
\centering
\includegraphics[width=0.75\textwidth]{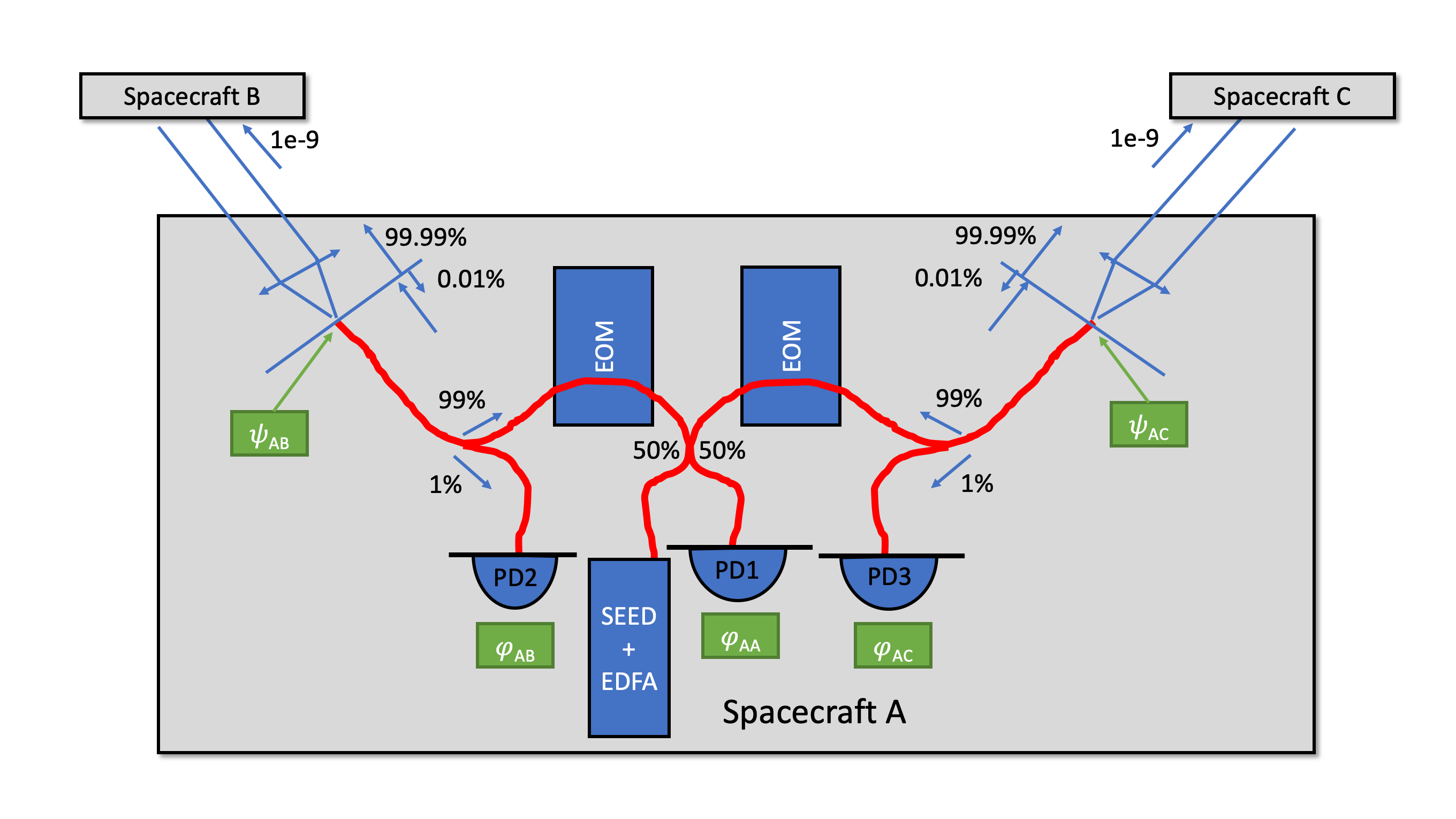}
\caption{ Representation of the optical setup for one of the spacecrafts. The red curves are single mode fibers. A seed laser  at $1.55\,\mu$m is amplified, split into two beams, phase modulated with two LiNbO3 electro-optic modulators, and collimated in the direction of the two other spacecrafts. The two beams are also weakly back-reflected into the fibers from the fiber extremity (where the beam leaves the fiber). At this point, the light also interferes with the incoming laser beams from spacecrafts 2 and 3. Photodiodes 2 and 3 measure the phase difference between the two outgoing beams and the two incoming beams. Last, diode D1 provides an internal metrology measurement which will monitor the non-common optical path inside the fibers.
\label{fig:schema}}
\end{figure*}

Fig~\ref{fig:schema} shows the optical layout of the interferometric bench of the satellites. For robustness and compactness, the SAGE optical bench is made at 100\% from fibered optics components. The spacecraft telescopes, described in Sec.~\ref{sec:telescopes}, are the only parts in bulk optics. The seed laser is a butterfly packaged external cavity laser (ECL) controlled in current and temperature. The beam is powered up at 200\,mW by the mean of an Erbium-Doped Fiber Amplifier (EDFA). Beam splitters are fibre couplers working from evanescent light, removing the risk of back reflection of the laser light. Time synchronisation between the satellites is obtained thanks to two LiNbO$_3$ electro-optic modulators (EOM). On the contrary to LISA, where the requirements is to have an absolute wavelength, the lasers of SAGE will be tuned together within the GHz bandpass of the photodiodes once the optical link is created. 

The basic principle of the metrology is time delayed heterodyne interferometry (TDI) \cite{2014LRR....17....6T} at $1.55\,\mu$m. After dividing the beam with a 50/50 fused fiber coupler, the  telescopes send two beams at 60 degrees from each other to the other two spacecrafts. For a spacecraft $A$, there are two reference optical positions: $\psi_{AB}$ and $\psi_{AC}$. These positions correspond to the two semi-reflective extremities of the two single mode fibers launching the beams into free space (see Fig.~\ref{fig:schema} for a representation of the optical layout inside the satellite). They are used as reference positions because they are the last flat optical surfaces before collimation towards, respectively, spacecrafts $B$ and $C$. The optical distance between the satellites is measured between these reference positions. The optical path measurement between satellite $A$ and $B$ is obtained between the reference positions $\psi_{AB}$ and $\psi_{BA}$. The ones between satellites $A$ and $C$ and $B$ and $C$ are respectively measured between positions $\psi_{AC}$ and $\psi_{CA}$, and $\psi_{BC}$ and $\psi_{CB}$. 

Fibers are very sensitive to temperature variations. Therefore an internal metrology measurement is necessary and is obtained within each satellite between $\psi_{AB}$ and $\psi_{AC}$ (inside satellite A), $\psi_{BC}$ and $\psi_{BA}$ (satellite B), and $\psi_{CA}$ and $\psi_{CB}$ (satellite C). Onboard each satellite, it is obtained by back-reflection of the fiber source on each fiber extremity.

\subsection{Time delay interferometry}

Within each spacecraft, the phases of the electromagnetic signal are 
measured from the heterodyne signal via GHz photodiodes. The photodiodes are
labeled $PD1$, $PD2$ and $PD3$ in Fig.~\ref{fig:schema}. The optical path length measured on $PD1$ corresponds to the internal metrology: 
 \begin{equation}
 \phi_{AA}(t)=2 \psi_{AB}(t)-2 \psi_{AC}(t)\ .
 \end{equation}
 The 
  $PD2$ and $PD3$ photodiodes measure the phase between the different satellites, respectively: 
 \begin{equation}
 \phi_{AB}(t)=\psi_{AB}(t)-\psi_{BA}(t-L_{BA})
 \end{equation}
 and
 \begin{equation}
\phi_{AC}(t)=\psi_{AC}(t)-\psi_{CA}(t-L_{CA})
 \end{equation}
 where the time delays $L_{BA}$ and $L_{CA}$ are necessary to account for the time of flight of the photons from satellite $B$ to $A$ and $C$ to $A$.

  The TDI measurement is a combination of all the optical path measurements such that the absolute optical phase values $\psi_{ij}$ become irrelevant. In the sensitivity calculation made in Section~\ref{sec:sensitivity}, we consider an unequal-arms-length Michelson configuration, a version of the Sagnac interferometer where the photons sweep through the interferometric arms 4 times. 
 These combinations are 8-pulse responses to gravitational waves called X, Y and Z\cite{1999PhRvD..59j2003T,2000PhRvD..62d2002E}. 
However, different configurations can also be used to optimise further the sensitivity at specific frequencies~\cite{2014LRR....17....6T}. In configuration X,
 the two interferometric arms consist in the 290\,000\,km trajectories passing by, in respective order, satellites $ABACA$ and $ACABA$. The final optical path difference (OPD) measurement is therefore (neglecting the internal metrology measurement) the difference between
 \begin{equation}
 \phi_{AB}(t)+\phi_{BA}(t-L_{AB}) +\phi_{AC}(t-L_{ABA})+\phi_{CA}(t-L_{ABAC}) 
 \label{eq:TDI}
 \end{equation}
and
 \begin{equation}
\phi_{AC}(t)+\phi_{CA}(t-L_{AC}) +\phi_{AB}(t-L_{ACA})+\phi_{BA}(t-L_{ACAB}) \,.
 \label{eq:TDI2}
 \end{equation}
Here, $L$ with multiple indices are the time of flight between several spacecrafts in respective order. For example,
 $L_{ACAB}$ corresponds to the time of flight of a photon going from satellite $A$ to $C$, back to $A$, and then to $B$ : $L_{ACAB}=L_{AC}+
 L_{CA}+L_{AB}$. The difference between (\ref{eq:TDI}) and (\ref{eq:TDI2}) gives a phase equivalent to the phase of a light beam going 
through the two interferometric optical paths corresponding to the $ABACA$ and $ACABA$ arms.

The 8-pulse response is needed because of the rotation of the antenna around the Earth. In geostationary orbit, the equilateral triangle moves with an angular speed of 360$^\circ$ per day. As a result, a standard clockwise and anticlockwise Sagnac measurement ($ABCA$ versus $ACBA$) would produce a differential optical path of the order of 1\,km. On the contrary, an important advantage of the X Sagnac configuration is that the optical path difference between the two arms of the interferometer is almost zero \cite{2014LRR....17....6T}: the photons travel in both direction in both arms. The residual OPD is then caused by a satellite drifting away from its geostationnary orbit: for a differential speed of the order of 0.7\,m/s \cite{2015CQGra..32r5017T}, the difference between the length of the two $ABACA$ and $ACABA$ arms is less than one meter, well within the coherence length of the laser. 

\subsection{Time synchronisation and ground telemetry}

A downside of the CubeSats is the limited power budget, and therefore the bandpass to transmit data to Earth. 
 Indeed, the photodiodes record the data at GHz rates. Thus the raw data would be too large to be downloaded. However, the useful data can be reduced to the 
 bandpass of the GW detector. Using the difference between Eq.~(\ref{eq:TDI}) and (\ref{eq:TDI2}), the gravitational signal comes from the combination of three terms :
 \begin{eqnarray}
s(t_A)&=& s_A(t_A)+s_B(t_{A\rightarrow B})-s_C(t_{A\rightarrow C})
\label{eq:stA}
 \end{eqnarray}
 where $t_A$ is the time of the clock onboard spacecraft $A$, $t_{A\rightarrow B}=t_A-L_{AB}$, $t_{A\rightarrow C}=t_A-L_{AC}$, and:
 \begin{eqnarray}
s_A(t_A)&=&  \phi_{AB}(t_A) -   \phi_{AB}(t_A-L_{ABA})  - \phi_{AC}(t_A) +  \phi_{AC}(t_A-L_{ACA})  \nonumber \\
s_B(t_{A\rightarrow B})&=& \phi_{BA}(t_{A\rightarrow B}) -  \phi_{BA}(t_{A\rightarrow B}-L_{ACA})   \nonumber \\
s_C(t_{A\rightarrow C})&=& \phi_{CA}(t_{A\rightarrow C}) -  \phi_{CA}(t_{A\rightarrow C}-L_{ABA})  
\label{eq:stA2}
 \end{eqnarray}
 Each signal is a phase measurement performed on a different spacecraft, respectively satellites $A$, $B$ and $C$.
In absence of acceleration, their linear combination is equal to zero: $s(t_A)=0$. 
 The data can be binned onboard each one of satellites if and only if: 
 \begin{enumerate}
 \item satellite $A$ has knowledge of the time of flight to satellite B and C ($L_{ABA}$ and $L_{ACA}$) ,
 \item
  satellite $B$ knows $L_{ACA}$ as well as the time $t_A$ onboard spacecraft $A$ including time of flight to $B$ ($t_{A\rightarrow B}$),
  \item satellite $C$ knows  $L_{ABA}$ as well as $t_{A\rightarrow C}$.
  \end{enumerate}

 The optical distances, $L_{ABA}$ and $L_{ACA}$, can be derived from the internal clocks of the respective spacecrafts:
 \begin{equation}
 L_{ABA}= (t_{B\rightarrow A}-t_A) - (t_{A\rightarrow B}-t_B)   \,. \label{eq:LABA}
 \end{equation}
The data necessary to time-synchronise the measurements onboard all satellites is therefore the three clocks
 $t_A$,  $t_B$,  $t_C$,  but also the six delayed clock values $t_{A\rightarrow B}$, $t_{B\rightarrow A}$, $t_{A\rightarrow C}$, $t_{C\rightarrow A}$, $t_{B\rightarrow C}$,  and $t_{C\rightarrow B}$.
For each value, synchronisation functions will be encoded in the side-bands of the metrology laser by the EOMs, and not send to ground. The only scientific telemetry send to ground will  be the 3 values corresponding to the 3 variables in the right hand of Eq.~(\ref{eq:stA}), binned at 5 Hz to respect Shannon's law, and then assembled on the ground to give the science grade dataset. For each spacecraft, the science baudrate can therefore be as low as 3 doubles at 5Hz, or 120 octets/s.

\section{Error terms}
\label{sec:error}

\subsection{Overall view}

Table~\ref{tb:noise} gives an overview of the error budget of the SAGE project with $N(f)$ the power spectrum density of the noise terms. The error budget is split into several categories. The quantum limitation is a hard limitation related to the number of photons received by the satellites. The external forces are hard limitations due to the absence of internal inertia measurement. The center of mass (CM) error is related to the difficulty to properly shield the center of gravity of the spacecraft with respect to the satellite itself. The interferometric measurement errors are related to the TDI measurement, and the scattered light is the error caused by light bouncing back into the fiber. The last source of errors comes from random variations of known gravitational fields: Moon, Sun, Earth as well as unresolved galactic binaries.

Each noise level is translated into a sensitivity limit assuming the main parameters of the mission from Table~\ref{tb:param}, in pm/$\sqrt{\rm Hz}$ relative to the 73\,000\,km length between two satellites. For example, the irradiation noise is calculated assuming a satellite cross section of 30\,cm$\times$20\,cm and a mass of 20\,kg.

\begin{table*}
   \caption{\label{tb:noise} 1$\sigma$ Error Budget}
   \centering
   \begin{tabular}{| l | l | c | c| c |} 
     \hline
     Category & Type & Eq. & Main parameter & $\sqrt{N(f)}$ [pm/$\sqrt{\rm Hz}$]  \\
     \hline
     \hline
     Quantum & Photon noise & (\ref{eq:photons}) & $P_{\rm avail}=15$\,pW & 23 \\
     External & Solar Irradiation & (\ref{eq:irradiation}) & 10\% calibration & $1.4\times10^{-6} f^{-4.25}$ \\
     External & Solar Wind & (\ref{eq:wind}) & 10\%  calibration & $3.3\times10^{-3} f^{-2.75}$ \\
     CM & Tilt-to-length & (\ref{eq:pointing}) & 0.05\,mas/$\sqrt{\mathrm{\rm Hz}}$ & 10 \\
     CM & Thermal expansion & (\ref{eq:thermal}) &  $T_{\rm stable}= 0.1\,$K  & $1\times f^{-1}$  \\
     Interferometric & Amplification noise & (\ref{eq:vacuum}) & $P_{\rm ampli}=0.0025$\,pW & 0.3 \\
     Interferometric & Flicker noise & (\ref{eq:clock}) & $v=1\,$m/s & $0.12\times f^{-0.5}$ \\
     Interferometric & Frequency noise & (\ref{eq:freq}) & $F_{\rm laser} = 10\,{\rm kHz}/\sqrt{\rm Hz}$ & 15.5 \\
     Scattering & laser light & (\ref{eq:scatter}) & $P_{\rm scattered}=0.02$\,pW & 2.9 \\
     \hline
   \end{tabular}
\end{table*}

\subsection{Quantum Noise}
\label{sec:measure}

The quantum noise, also called shot-noise, comes from the quantum nature of light: each photon as an arrival uncertainty of $\approx 1$\,radian. Only by increasing the number of photons can one decrease that source of noise. As a result, the more powerful the laser is, the better it is. However, the  limited power budget and the thermal stability of the spacecraft requires a moderate laser energy consumption.

To be in line with a 12U CubeSat (with typical power budget of the order a 20\,W) we considered a 200\,mW single frequency fibered laser diode (2\,W power consumption). The diffraction caused by the pupil and the 73\,000\,km distance of the satellite is showed in  Fig.~\ref{fig:diff}.
This large pupil shadowing is due to the fact that the two telescopes are intertwined within a limited space. The resulting diffraction, in the Fraunhoffer approximation, is showed in the right panel of the same figure. It causes the amplitude of the beam to be fainter by $10^{-7}/\text{m}^2$ from the total emitted light. 

The coupling efficiency out and into the fiber can also be calculated and amount to 50\% for each~\cite{2018arXiv180608106L}. In the end, from the 100\,mW of optical power emitted by each fiber, only $100\,$mW$\times 1.5 \times 10^{-10}=15\,$pW is received by the fiber on the other spacecraft.
At a wavelength $\lambda=1.55\, \mu$m, the accuracy of the metrology measurement is then limited by the photon noise:
 \begin{equation}
 \sqrt{N_{\rm laser}(f)}=\sqrt{\frac{ \hbar c\lambda}{2\pi P_{\rm avail}}}=23\,{\rm pm}/\sqrt{\mathrm{\rm Hz}}
 \label{eq:photons}
 \end{equation}
 
\begin{figure*}
\centering
\includegraphics[width=0.9\textwidth]{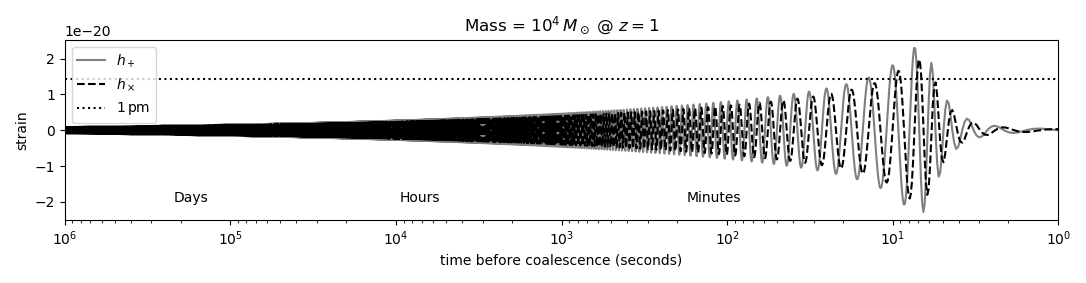}
\caption{ Inspiral amplitude of the two polarized signals for the coalescence of two IMBH of masses $10^4\,M_\odot$, at $z=1$. From PyCBC software.
\label{fig:inspi}}
\end{figure*}

 The strain of the most energetic gravitational waves is of the order $10^{-21}$. For an interferometer of arm length 73\,000\,km, the effect of the gravitational wave is 0.7\,pm in amplitude. It seems small with respect to $23\,\mathrm{pm}/\sqrt{\mathrm{\rm Hz}}$. However, the baseline concept is to fit inspiral patterns over a month of observations. In the example case of Fig~\ref{fig:inspi}, where two IMBH merge at a distance of $z=1$, the strain of amplitude $10^{-21}$ can be detected with a signal to noise ratio of 8 in a day of science operation (assuming photon noise only).

\subsection{Solar irradiation and solar wind}
\label{sec:extern}

The acceleration caused by solar irradiation is of the order $10^4$ higher than the signal expected from a GW. However, most of the energy is at low frequency: the total solar irradiance (TSI) changes slowly with time. To characterize its power spectral density (PSD), we used data from the VIRGO (Variability of solar IRradiance and Gravity Oscillations) instrument on-board the SOHO observatory~\cite{Frohlich95,Frohlich97,2002SoPh..209..247J}.

\begin{figure*}
\centering
\includegraphics[width=0.8\textwidth]{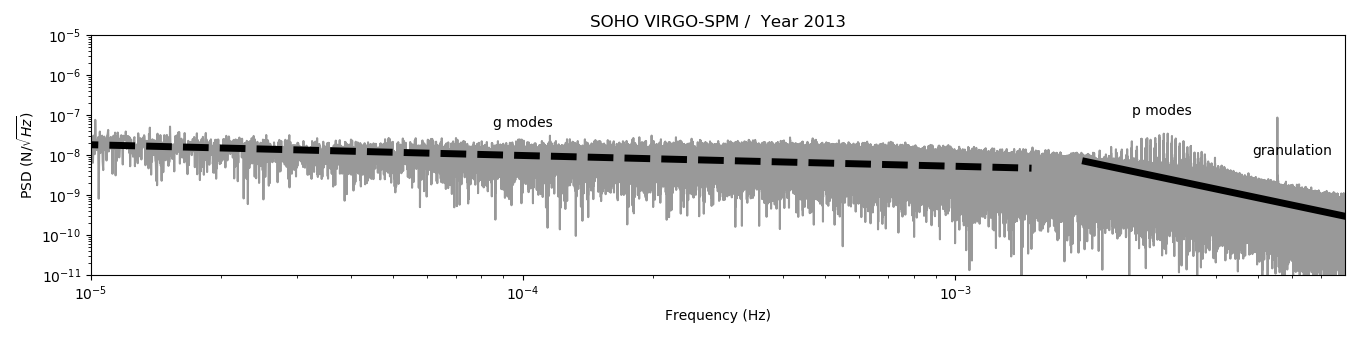}
\includegraphics[width=0.8\textwidth]{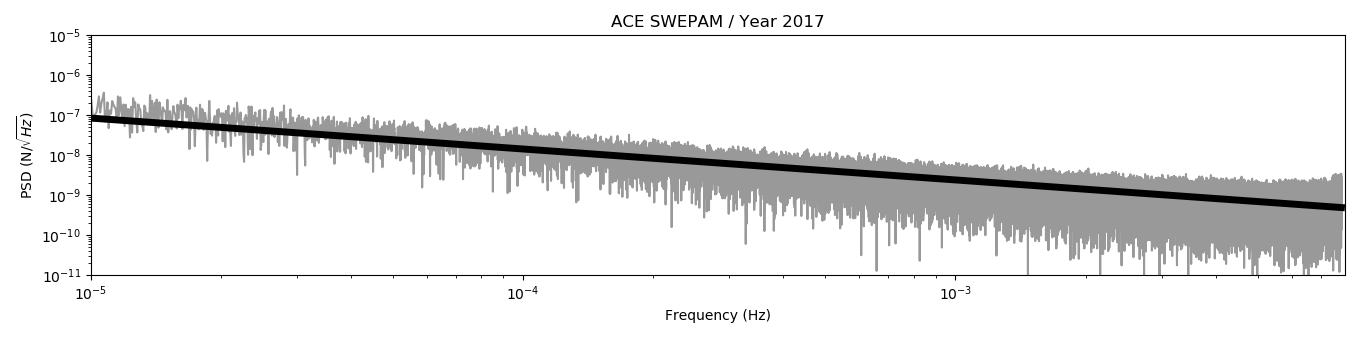}
\caption{ {\bf Upper panel -- Solar Photons:} power spectrum density of the solar irradiation pressure calculated from the green channel of the SPM instrument. The dashed curve is the fit of the gravitational modes, and the solid curve is the fit of the granulation. The 3\,mHz forest lines ($p$-modes) are excluded from the fit.
{\bf Lower panel -- Solar Protons:} Power spectral density of the solar wind pressure during the period 2017 from ions density and speed measured by the SWEPAM instrument aboard the ACE observatory.
\label{fig:photons}}
\end{figure*}

In Fig.~\ref{fig:photons} we present the PSD from a dataset that covers the full year 2013. One can clearly see the pressure modes that dominate the PSD around 3\,mHz (resonance of the radiative structure). Below 1\,mHz, the spectrum is dominated by the gravitational modes (convective structure). The trend, below and after the $p$-modes, are caused by the granulation. Data sampling is 60\,s, which means that the maximum frequency (Shannon's frequency) is 8\,mHz. However, it is possible to extrapolate the PSD at higher frequency. We have fitted the granulation noise by a logarithmic relation: $\sqrt{N(f)}=\alpha \times f^\mu$. It yields:
\begin{equation}
\sqrt{N_{\rm irradiation}(f)}=4.8\times10^{-15} f^{-2.25} \,\mathrm{N}/\mathrm{m}^2/\sqrt{\rm \mathrm{\rm Hz}}
 \label{eq:irradiation}
\end{equation}
At the same time, the shot noise can be deduced from the ratio between total energy ($P_{\rm photon}=1000 \,$W/m$^2$) and mean energy of a $600$\,nm photon. It yields to a negligible value compare to the granulation noise.

Similarly, we estimated the acceleration caused by the solar wind. We restricted the analysis to the low speed protons ($v\ll c$) which are emitted from the Sun during periods of solar activity. This wind is monitored from the L1 Lagrange point by the SWEPAM instrument on board the ACE satellite. The instrument monitor both the speed $v$ and the density $d$ of the ions in the solar wind. The force applied to the satellite, in N/m$^2$, is calculated from the mass of a proton $m_p$: $F_{\rm protons}=m_p*d*v^2$. In the lower panel of Fig.~\ref{fig:photons} is plotted power density of this force. The fit to the PSD gives:
\begin{equation}
\sqrt{N_{\rm wind}(f)}=1.1\times10^{-11} f^{-0.744} \,\mathrm{N}/\mathrm{m}^2/\sqrt{\rm \mathrm{\rm Hz}}\,.
 \label{eq:wind}
\end{equation}
The shot noise can also be calculated and is negligible, close to $10^{-16}  \,\mathrm{N}/\mathrm{m}^2/\sqrt{ \mathrm{\rm Hz}}$ \cite{2018arXiv180608106L}. On the contrary to the L1 Lagrange point, the solar wind at geostationary altitude can have his orientation changed by the upper Van Allen belt. This can be monitored by putting individual sensors on each spacecraft.

\subsection{Center of mass stability}

The satellite center of mass (CM) will orbit around the Earth following a ballistic trajectory, with the exception of external forces stated in section~\ref{sec:extern}. But the OPD measurement is not exactly made between the spacecrafts. It is made between the two fibers extremities that are on two different spacecraft (Figure~\ref{fig:schema}), and several effects can modify the position of the fibers extremities with respect to the center of mass. The first is a rotation of the spacecraft around the CM. The second is a mechanical distortion of the spacecraft due to thermal expansion.

\subsubsection{Tilt-to-length coupling}

The OPD measurement is done between the two reference positions which are  the extremities of the fibers ($\psi_{AB}$ and $\psi_{AC}$ in Fig.~\ref{fig:schema}). Even if the center of mass is stable, the position of the two reference positions may move with respect to it. The simplest example is a rotation of the satellite. Fortunately, this rotation can be measured and post-processed thanks to our knowledge of the optical orientation with respect to the two other spacecrafts. 

The orientation of the satellites is obtained by the location of the incoming laser light in the focal planes of the telescopes. This knowledge is obtained by maximisation of the injection into the fiber, and incidentally, by the measurement of the position of the fibers in the focal plane. This measurement is obtained by the sensor gauges (SG) which are typically capacitive or resistive sensor. Assuming a focal length of the optical system of 40\,cm and an accuracy of the SG of 0.1\,$\mu$m/$\sqrt{\mathrm{\rm Hz}}$, it means that the orientation of the spacecraft can be estimated (with respect to the other spacecrats) below 0.05\,''/$\sqrt{\mathrm{\rm Hz}}$ (the point spread function size is 3.5'').

The effect on the OPD will depend on the level of collinearity between the  optical system and the position of the CM. We consider the collinearity level at 0.002 radians (correspond to the pointing accuracy of the satellites). We also consider a distance between the CM and the optimum position where the tilt-to-lenght coupling is zero to be $l=4$\,cm in virtual space. Then,
 the displacement of the reference position with respect to the center of mass can be determined with a precision of $0.002\times4\,/2$cm times the pointing precision:
\begin{equation}
\sqrt{N_{\textrm{tilt-to-length}}(f)}= 10\, \mathrm{pm}/\sqrt{\rm \mathrm{\rm Hz}}\,.
\label{eq:pointing}
\end{equation}

\subsubsection{Thermal expansion}

Another problem comes from the deformations caused by the thermal expansion of the spacecraft. Even if the full CubeSat is made of a material with a very low thermal expansion coefficient (eg, SiC). For example, with a material of thermal expansion of $\alpha=2\times 10^{-6}$\,K$^{-1}$, and considering the typical size of 10\,cm, a temperature gradient in the spacecraft would move the CM by 200\,nm/K$^{-1}$. However, the heat capacity of the spacecraft damps the temperature variations. A thermal system can be seen as a first order band pass filter, with a characteristic cutoff frequency, and a characteristic amplitude.

For a satellite of 20\,kg, the heat capacity of the spacecraft is of the order of $c_{\rm th}=10^4$~J/K. Assuming a dissipated heat of $P_{\rm dissipated}=10\,$W which transfers within a system already passively stabilised within $T_{\rm stable}=0.1\,$K, it means that the characteristic timescale for temperature variation is $\tau_{\rm stable}=c_{\rm th}T_{\rm stable}/P_{\rm dissipated}=10^4\,$s. The characteristic amplitude, assuming $P_{\rm white}=5\,$W  as a white noise energy dissipation, is then ${T_{\rm stable}}{P_{\rm white}}/{P_{\rm dissipated}}$.

From the characteristic amplitude and cutoff frequency, the power spectrum of the CM displacement caused by thermal expansion can be derived. For $f\gg1/\tau_{\rm stable}$, it writes:
\begin{equation}
\sqrt{N_{\rm thermal}(f)}=\frac{P_{\rm white}}{P_{\rm dissipated}} \frac{T_{\rm stable}}{f\tau_{\rm stable}}  \alpha \times 10\,\mathrm{cm}=1\,f^{-1}\,\mathrm{pm}/\sqrt{\mathrm{\rm Hz}}\,.
\label{eq:thermal}
\end{equation}

We do not foresee operation when the satellites can be eclipsed by the Earth.  The thermal stress on the
spacecrafts and the associated relaxation processes would impair the observation. However, the equatorial plane is inclined by $23.5^\circ$ with respect to the orbital plane of the Earth around the Sun. It means that these eclipses only happen around equinox, twice a year, for a period of $\approx 20\,$days.

 \subsection{Interferometric Noise}

 \subsubsection{Amplification or Vacuum noise}
 
The fundamental principle of TDI is that it relies on heterodyne interferometry. The problem with heterodyne interferometry (which limits its application to radio wavelength in astronomy), is that it adds a noise which is inversely proportional to the number of photons per coherence time. Using an external cavity diode laser of sufficiently large cavity size ($\approx 8\,$mm), the emission linewidth can be narrowed down to 20\,kHz. It means an energy for the vaccum noise emission on the amplified signal of $P_{\rm ampli}=20\times10^3 h\nu=0.0025\,\mathrm{pW}$. Compared to the energy received of $P_{\rm avail}=15\,\mathrm{pW}$, the noise on the OPD measurement is:

 \begin{equation}
\sqrt{N_{\rm ampli}(f)}= \sqrt{\frac{\hbar c\lambda}{2\pi }}\times\frac{\sqrt{P_{\rm ampli}}}{P_{\rm avail}}=0.3\,{\rm pm}/\sqrt{\mathrm{Hz}} \,.
\label{eq:vacuum}
 \end{equation}
  
 \subsubsection{Clock noise}

The proposed 8-pulse TDI measurement is spelled in Eqs.~(\ref{eq:stA}) and (\ref{eq:stA2}). The signal is the linear combination of three phase measurements made on the three spacecrafts. The noise on this measurement comes from a combination of heterodyne beating frequency and clock noise. The higher the heterodyne frequency, the more accurate the clock must be synchronised between the different spacecrafts.

%
%For example, on satellite $C$, the measurement $s_C(t_{A\rightarrow C})$ is 
%\begin{equation}
%s_C(t_{A\rightarrow C})=\phi_{CA}(t_{A\rightarrow C}) -  \phi_{CA}(t_{A\rightarrow C}+ (t_{A\rightarrow B}-t_B)_{\rightarrow C} -  (t_{B\rightarrow A}-t_A)_{\rightarrow C}  )\,,
%\end{equation} 
%where $t_{A\rightarrow C}$ is the time onboard satellite $A$ delayed to satellite $C$, and $L_{ABA}$ is twice the optical distance between satellites $A$ and $B$. The noise on this measurement comes from a combination of heterodyne beating frequency and clock noise.
%
%
%
%of the two lasers. This beating frequency can as 
%
%
%The noise caused by the clock uncertainties in the TDI measurement depends on own well are known $t_{A\rightarrow C}$ and $L_{ABA}$, but also how fast $\phi_{CA}(t)$ changes. By tuning the heterodyne frequency, we
%
%our capacity to synchronise the spacecrafts and control the phase variation in each spacecraft. The slower $\phi$, the more relax the constrain on the time syncrhonisation of the spacecraft.
%
%cancel the phase error as showed by Eq.~(\ref{eq:TDI}) and (\ref{eq:TDI2}). The linear combination perfectly cancel the phase noise if the spacecrafts have either i) a perfect common time reference, or ii) a phase $\phi(t)$ independent of time. With moving spacecrafts, $\phi(t)$ changes, and hence, the satellites must synchronise their onboard clocks.
The minimum heterodyne frequency is determined by the relative speed of the spacecrafts.
In the paper by \cite{2015CQGra..32r5017T}, it is shown that the relative speed between the spacecrafts can be maintained below $v=0.7$\,m/s during a week without station keeping. According to the same paper, we can write the effect on the power spectrum of the "flickering clock noise" as :
\begin{equation}
\sqrt{N_{\rm TDI}(f)}= \frac{\sigma_A v}{\sqrt{2\ln2}c} f^{-0.5}\,{\rm strain}/\sqrt{\rm Hz}=1.7\times10^{-21} f^{-0.5}\,{\rm strain}/\sqrt{\rm Hz}\,,
\label{eq:clock}
 \end{equation}
 where $\sigma_A$ is the stability of the onboard clock which is typically characterised by the Allen deviation (ADEV). Here, we have taken $\sigma_A =10^{-11}$ at 1\,Hz, which is available for off the shelf rubidium oscillator sources.
 
 \subsubsection{Phase noise}

 Synchronisation between the spacecrafts is done by the same beam that is used for the metrology. The time signal will be coded in phase offsets generated by the LiNbO3 modulators. This is equivalent of using the time of flight of the photons to determine the optical distance between the spacecrafts. We expect a synchronisation at the GHz level which corresponds to the bandpass of the photodiode. 
 This mean that the errors on the time variables in Eq.~(\ref{eq:stA2}) will be of $\sigma_t=1\,$ns,
 This is equivalent to knowing the separation between the satellites at $c \times \sigma_t=0.3\,$m.
 For a laser  with a white frequency noise of $F_{\rm laser} = 10\,{\rm kHz}/\sqrt{\rm Hz}$, it means that the phase noise coming the change of frequency of the laser  will be:
\begin{equation}
\sqrt{N_{\rm frequency}(f)}= \lambda \sigma_t  F_{\rm laser}=15.5\,\mathrm{pm}/\sqrt{\mathrm{Hz}}\,.
\label{eq:freq}
 \end{equation}
 For this level of phase noise, no amplitude stabilisation is required: the relative intensity noise does not dominate the phase error. The relatively low requirement on the stability of the laser is obtained  thanks to the accurate clock synchronisation between the phase measurements, only possible with fast photodiodes.

 \subsubsection{Scattered light noise}

Thanks to the use of single mode fibers, the only scattered light that goes back into the interferometer is the light scattered by the optical elements inside the numerical aperture of the fiber. The most sensitive optical part is the M2. The light back reflected through scattering toward the fibers can be 0.1\%. At a distance $d_1=10\,$cm from the $d_2=5\,\mu$m fiber core, the solid angle of the fiber core is $\pi (d_2/2d_1)^2$. The flux of
 back scattered light is therefore:
\begin{equation}
P_{\rm scattered}=0.1\%\frac{200\, \mathrm{mW}}{2}\frac{\pi (d_2/2d_1)^2} {4\pi}=0.02\,\mathrm{pW} \,.
\end{equation}
This scattered light adds up to the received energy from the other satellite. It results in a phase error that we can assume to be random with an amplitude of ${P_{\rm scattered}}/{P_{\rm avail}}$ and a timescale which corresponds to the coherence time of the laser (1/50\,kHz). Thus, the power spectrum of the noise caused by scattered light writes:
\begin{equation}
\sqrt{N_{\rm scattering}(f)}= \frac{\lambda}{2\pi} \frac{P_{\rm scattered}}{P_{\rm avail}} \frac{1}{\sqrt{50\,\mathrm{kHz}}}=2.9\,\mathrm{pm}/\sqrt{\mathrm{Hz}}\,.
\label{eq:scatter}
 \end{equation}
 
 \subsection{Earth, Moon, Sun, and the Galactic confusion noise}

A clear advantage of the geostationary orbit is to have the antenna fix with respect to the Earth rotation. This eliminate large portions of the noise caused by the heterogeneous Earth gravitational field. But Moon, Sun, and the other planets create a noise with a period of approximately 24 hours. The randomness of this noise is out of the scope of this paper, but can open a wide field of research by themselves. The static components on the 24 hours period have to be taken into account during processing.

The signal from galactic binaries is also something that the experiment cannot avoid. As the mission duration extend, the main source of noise can be determined and can also be removed during data reduction. However the frequency of this noise matters more for LISA than from SAGE, with typical frequency of the order of a few mHz and a strain amplitude below $10^{-18}/\sqrt{\rm Hz}$ for a 4 years mission duration \cite{2017JPhCS.840a2024C,2018arXiv180301944C}. We designed the SAGE mission for 3 years of operation, but a long term observatory can be envisioned if we accept to replace the satellites in case of failure.

%%% new

\section{Sensitivity}
\label{sec:sensitivity}

\subsection{Optical path delay measured by SAGE}

\begin{figure}[htbp]
\centering
\includegraphics[width=0.47\textwidth]{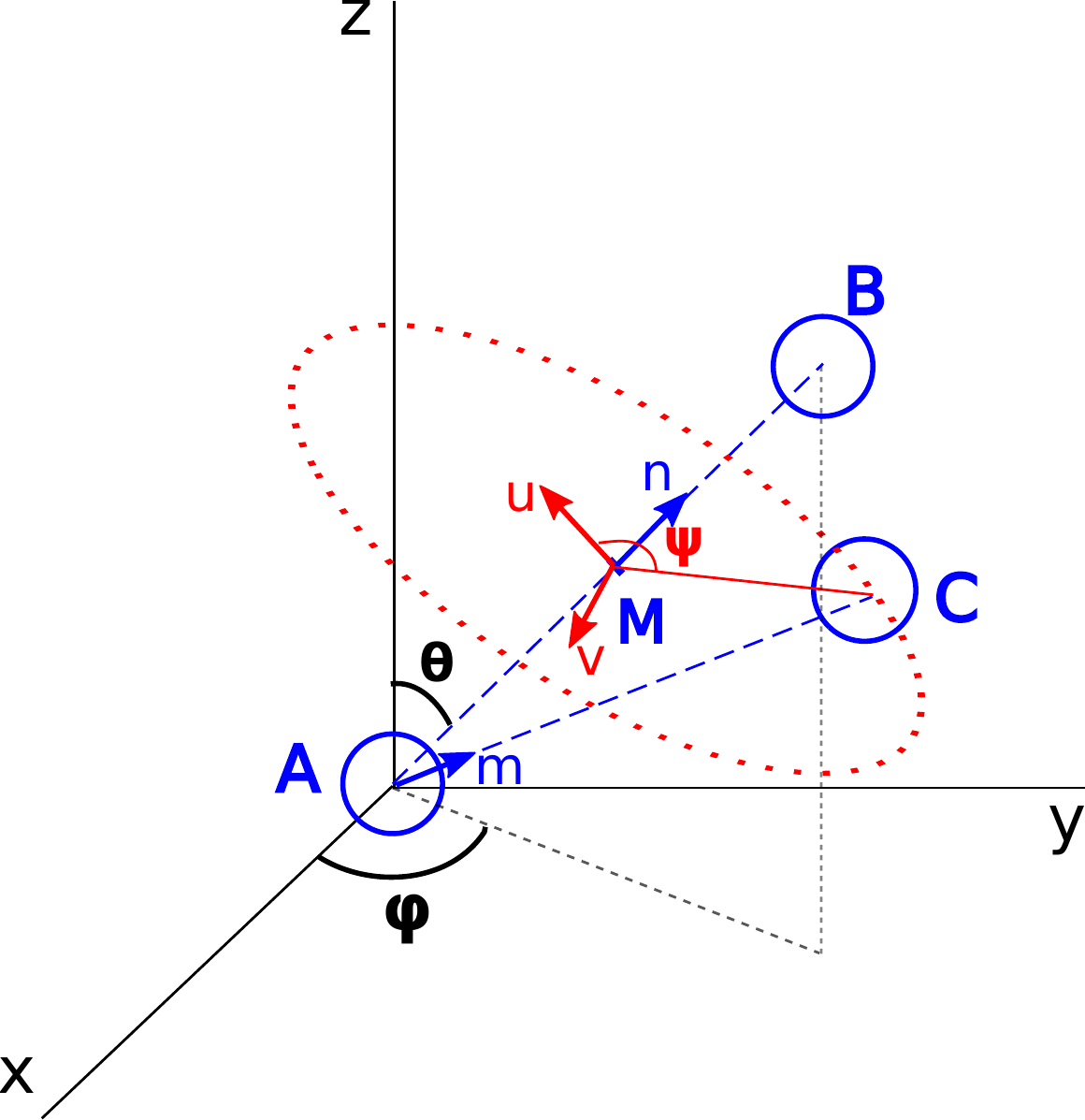}
\caption{Geometry of the problem. See text for details.}
\label{fig:geom}
\end{figure}

Let us consider the geometry of Fig.~\ref{fig:geom}.
We have three spacecrafts, A, B, and C, forming an equilateral
triangle with arm length $L$. Let us consider a plane gravitational wave
with wave vector pointing along the direction $z$ at spacecraft $A$. 
This direction is completed to form a direct triad $(x,y,z)$ in which all the
computations will be derived. Spacecraft B is labeled by its spherical
coordinates $(L,\theta,\pp)$ in this triad. Let 
\be
\mathbf{n} = (\sin\theta \cos\pp,\sin\theta \sin\pp,\cos \theta)
\ee
be the unit
3-vector along AB. Let M be the middle of the AB segment. Spacecraft C
lies in the plane orthogonal to $\mathbf{n}$ passing through M (hereafter, the "mid plane"), and more
precisely on a circle centered on M with radius $\sqrt{3}/2\,L$, depicted
in red dotted in the figure. We want to label the position of spacecraft C
in the mid plane. To do so, we consider a direct triad $(\mathbf{u},\mathbf{v},\mathbf{n})$
such that $\mathbf{u}$ is the projection of $\mathbf{e_z}$ in the mid plane, perpendicular
to $\mathbf{n}$, and $\mathbf{v}$ completes the direct triad.
It is easy to get
\bea
\mathbf{u} &=& \frac{\mathbf{e_z} - \cos \theta \,\mathbf{n}}{\sin\theta} \\ \nn
&=& (-\cos\theta \cos \pp, - \cos \theta \sin \pp, \sin\theta). \\ \nn
\eea
Then, requiring that $\mathbf{v}$ is perpendicular to $\mathbf{u}$ and $\mathbf{n}$,
and that $\mathbf{u} \times \mathbf{v} = \mathbf{n}$, uniquely defines $\mathbf{v}$
such that
\be
\mathbf{v} = (\sin\pp, -\cos \pp, 0).
\ee
The unit vector $\mathbf{w}$ along the $\mathbf{MC}$ segment now reads
\be
\mathbf{w} = \cos \psi\,\mathbf{u} + \sin \psi \, \mathbf{v}
\ee
where $\psi$ is the angle between $\mathbf{u}$ and $\mathbf{MC}$,
positive in the sense imposed by the direct triad $(\mathbf{u},\mathbf{v},\mathbf{n})$,
see Fig.~\ref{fig:geom}.
Let us call $\mathbf{m}$ the unit vector along the AC arm.
We have
\bea
\mathbf{AC} &=& \mathbf{AM} + \mathbf{MC} \\ \nn
L \, \mathbf{m} &=& \frac{L}{2} \mathbf{n} + \frac{\sqrt{3}}{2} L \mathbf{w} \\ \nn
\eea
so that
\bea
\mathbf{m} = &&\left(\frac{1}{2}\sin\theta \cos\pp - \frac{\sqrt{3}}{2} \cos\psi \cos \theta \cos \pp + \frac{\sqrt{3}}{2} \sin\psi \sin \pp, \right. \\ \nn
 && \frac{1}{2}\sin\theta \sin\pp - \frac{\sqrt{3}}{2} \cos\psi \cos \theta \sin \pp - \frac{\sqrt{3}}{2} \sin\psi \cos \pp, \\ \nn
 && \left. \frac{1}{2}\cos\theta + \frac{\sqrt{3}}{2} \cos\psi \sin \theta\right). \\ \nn
\eea

We now want to compute the time it takes for a photon to follow the
track $A-B-A-C-A$ along the arms of the triangle, taking into account
the effect of the gravitational wave. So we switch to a relativistic style.
We consider the background spacetime to be that of Minkowski, described
in Cartesian coordinates by the metric
\be
\eta_{\mu\nu} = \mathrm{diag}(-1,1,1,1).
\ee
This means that the spacetime interval between two neighboring
spacetime events reads
\be
\dd s^2 = \eta_{\mu\nu} \,\dd x^\mu \dd x^\nu = -\dd t^2 + \dd x^2 + \dd y^2 + \dd z^2
\ee
where we use the Einstein convention of summing over repeated indices, 
we keep the $(x,y,z)$ space coordinates introduced above,
and introduce a coordinate time $t$, which coincides with the proper
time of a static observer in the absence of gravitational wave.
In the expression above, $x^\mu$ is a general coordinate
that in our particular problem encapsulates $(x^0 = t, x^1 =x, x^2=y, x^3 = z)$.
A weak planar gravitational wave is now present, propagating along
direction $z$. It adds a small perturbation to the Minkowski metric
\be
h_{\mu\nu} = \begin{pmatrix}
0 & \mathbf{0} \\
\mathbf{0} & h_{ij}
\end{pmatrix}, 
\quad \mathrm{with} \:\:h_{ij} = \begin{pmatrix}
h_+ & h_\times & 0 \\
h_\times & -h_+ & 0 \\
0 & 0 & 0 \\
\end{pmatrix}
\ee
where we use the standard convention that Greek indices
run over the spacetime dimensions $0,1,2,3$, while Latin indices
run over the spatial dimensions $1,2,3$. Note that $h_{\mu\nu}$
is traceless, due to a particular choice of gauge that is made
to simplify the computations (the so-called "transverse-traceless",
or TT gauge).
The quantities $h_+$
and $h_\times$ encapsulate the two polarizations of the transverse
gravitational wave. They read
\be
h_{+,\times} (t,z) = h_{+,\times}(t - z)
\ee
where we use a system
of units where the velocity of light $c=1$.
The wave being weak,
we have
\be
h_{\mu\nu} \ll \eta_{\mu\nu}.
\ee
The full spacetime interval now reads
\be
\dd s^2 = \left( \eta_{\mu\nu} + h_{\mu\nu} \right) \, \dd x^\mu \dd x^\nu.
\ee

Let us consider a photon propagating between spacecrafts A and B.
It follows a null geodesic, so that $\dd s^2 = 0$ over its journey.
Let us consider two neighboring spacetime events, $E_1 = (t,x,y,z)$
and $E_2 = (t+\dd t, x+\dd x,y+\dd y, z+\dd z)$, occupied by the photon
at two infinitely close coordinate times along its trajectory from A to B. 
The spatial interval reads
\be
(\dd x^1, \dd x^2, \dd x^3) = (\dd x, \dd y, \dd z) = \mathbf{n} \,\dd l
\ee
where we introduce the element of length along the arm, $\dd l$.
We thus have: $\dd x^i = n^i\,\dd l$.
The spacetime interval between $E_1$ and $E_2$ thus reads
\bea
\dd s^2 &=& g_{tt} \, \dd t^2 + g_{ij} \, \dd x^i \dd x^j \\ \nn
&=& - \dd t^2 + (\eta_{ij} + h_{ij}) \,n^i n^j \, \dd l^2 \\ \nn
&=& 0 \\ \nn
\eea
where the last equality comes from the fact that the photon
follows a null geodesic of the perturbed spacetime.
The element of coordinate time between $E_1$ and $E_2$
is thus given by
\be
\dd t =  \sqrt{(\eta_{ij} + h_{ij}) \, n^i n^j}\, \dd l.
\ee
Let us expand this expression and write it to first order
in the metric perturbation:
\bea
\label{eq:dt}
\dd t &=&  \sqrt{(1+h_{xx}) \left(n^x \right)^2+(1+h_{yy}) \left(n^y \right)^2 + \left( n^z \right)^2 + 2 h_{xy} n^x n^y}\, \dd l \\ \nn
 &=&  \sqrt{1+ h_+\left[\left(n^x \right)^2- \left(n^y \right)^2\right] + 2 h_{\times} n^x n^y}\, \dd l \\ \nn
 &=&  \left(1+ \frac{1}{2} h_+\left[\left(n^x \right)^2- \left(n^y \right)^2\right] + h_{\times} n^x n^y\right)\, \dd l. \\ \nn
%&=& \sqrt{1 + h_+ \sin^2\theta \cos (2\pp) + 2 h_\times \sin^2\theta \cos \pp \sin \pp}\, \dd l \\ \nn
%&=& \left( 1 + \frac{h_+}{2} \sin^2\theta \cos (2\pp) + h_\times \sin^2\theta \cos \pp \sin \pp \right) \dd l. \\ \nn
\eea
A similar expression will be found when the photon moves along the $\mathbf{m}$
arm. Also, given that only products $n^i n^j$ appear, this expression is valid
for both ways of propagation (BA or AB, e.g.). We thus introduce the following
notation to encapsulate the directional information
\bea
\label{eq:ABCD}
\mathcal{A} &=& \frac{1}{2}\left[(n^x)^2 - (n^y)^2\right] =  \frac{1}{2}\sin^2\theta \cos (2\pp), \\ \nn
\mathcal{B} &=& n^x n^y = \sin^2\theta \cos \pp \sin \pp,\\ \nn
\mathcal{C} &=& \frac{1}{2}\left[(m^x)^2 - (m^y)^2\right], \\ \nn
\mathcal{D} &=& m^x m^y \\ \nn
\eea
where the two last expressions do not easily simplify.

Integrating Eq.~(\ref{eq:dt}) from $l=0$ to $l=L$
gives the travel coordinate time 
\be
t_1 - t_0 = L + \mathcal{A} \int_0^L h_+ \dd l + \mathcal{B}\int_0^L h_\times \dd l
\ee
where the coordinate time is $t_0$ when the photon leaves spacecraft A,
and $t_1$ when it reaches B.
We want to express the integrals as coordinate-time integrals to allow a Fourier
% treatment. So we will make the order-0 approximation $\dd t \approx \dd l$,
so that $l = t-t_0$ ($l=0$ at $t=t_0$ in A, $l=L$ at $t=t_1$ in B),
in the two integrals, and use $z = l \,n^z \approx (t-t_0)n^z $, to get
\be
\label{eq:baseq}
\int_0^L\, h_{+,\times} \dd l \approx \int_{t_0}^{t_1} h_{+,\times}(t-n^zt+ n^zt_0)\dd t.
\ee
We will further simplify the equation by using the top hat operator $\Pi_{t_0,t_1}$ and the convolution operator $\circledast$: 
\begin{equation}
\Pi_{t_0,t_1}(t)=\begin{cases}
1, \text{if }\ t_0<t<t_1 \\
0, \text{otherwise}
\end{cases}
\end{equation}
so
\bea
\int_{t_0}^{t_1} h_{+,\times}(t-n^zt+ n^zt_0)\dd t &=& \frac{1}{1-n^z} \int_{t_0}^{t_1-n^z(t_1-t_0)} h_{+,\times}(t)\dd t \\ \nn
&=&\frac{1}{1-n^z}  \int_{-\infty}^{+\infty} \Pi_{0,(n^z-1)(t_1-t_0)}(t_0-t)  h_{+,\times}(t) \dd t \\ \nn
&=&\frac{1}{1-n^z} \left[ \Pi_{0,(n^z-1)(t_1-t_0)}(t_0) \circledast h_{+,\times}(t_0) \right]
\eea
The final expression of the travel time difference is thus 
\bea
t_1 - t_0 = L &+& \left[ \frac{ \Pi_{0,(n^z-1)(t_1-t_0)}}{1-n^z}\right] \circledast \left[ \mathcal{A}h_{+}(t_0)+ \mathcal{B} h_{\times}(t_0)\right] \\ \nn
\eea
which depends on $t_0$ and $t_1$ the times when the photon quit and enter the spacecrafts.

The same computation can be done for the return trip BA. 
The basic equation is exactly the same
\be
t_2 - t_1 = L + \mathcal{A} \int_0^L h_+ \dd l + \mathcal{B}\int_0^L h_\times \dd l
\ee
where $t_2$ is the coordinate time when the photon is back at A, and
now $l=0$ at B and $l=L$ at A.
However there a modification when switching from $l$ to $t$ integration.
Indeed, now, $z = (L - l) \cos \theta$ to ensure that $z=z_B=L n^z$
at B where $l=0$. Moreover, $l = t - t_1$ to ensure that $t=t_1$ at B.
So the new expression of $z$ as a function of $t$ reads
$z = (L-(t-t_1)) n^z$. Eq.~(\ref{eq:baseq}) transforms to
\bea
\int_0^L\, h_{+,\times} \dd l &\approx& \int_{t_1}^{t_2} h_{+,\times}\left(t\,(1 + n^z) - (L+t_1)n^z\right)\dd t\\
					&=&  \frac{1}{1+n^z}  \int_{t_1-n^z(t_1-t_0)}^{t_2} h_{+,\times}\left(t\right)\dd t\\
					&=&\frac{1}{1+n^z} \left[ \Pi_{(n^z-1)(t_1-t_0),t_2-t_0}(t_0) \circledast h_{+,\times}(t_0) \right]
\eea
which  gives the travel time difference
\bea
t_2 - t_1 = L &+  &\left[ \frac{ \Pi_{(n^z-1)(t_1-t_0),t_2-t_0}(t_0) }{1+n^z}\right] \circledast \left[ \mathcal{A}h_{+}(t_0)+ \mathcal{B} h_{\times}(t_0)\right] \\ \nn
\eea

Let us express the relation $z(l)$ along the AC travel, with $l=0$ at $t=t_2$ in A and $l=L$ at $t=t_3$ in C. The altitude of C is
$z_C = m^zL$. In a similar fashion to the calculation over trajectory AB, the travel time along AC thus reads
\bea
t_3 - t_2 = L &+& \left[ \frac{ \Pi_{0,(m^z-1)(t_3-t_2)}(t_2) }{1-m^z}\right] \circledast \left[ \mathcal{C}h_{+}(t_2)+ \mathcal{D} h_{\times}(t_2)\right] \\ \nn
\eea

Along the return trip CA, $z = (L-l) \, C $, with $l = t-t_3$ where $t_4$ is the coordinate time back at A.
The CA travel time thus reads
\bea
t_4 - t_3 = L &+  &\left[ \frac{ \Pi_{(m^z-1)(t_3-t_2),t_4-t_2}(t_2) }{1+m^z}\right] \circledast \left[ \mathcal{C}h_{+}(t_2)+ \mathcal{D} h_{\times}(t_2)\right] \\ \nn
\eea

The ABACA complete travel time can be expressed with a not too heavy expression
provided that we use the convolution operator. The total travel time then reads
\bea
t_4 - t_0 = &4L&       \\ \nn
&+& \left[   \frac{ \Pi_{0,(n^z-1)(t_1-t_0)}(t_0) }{1-n^z}.  + \frac{ \Pi_{(n^z-1)(t_1-t_0),t_2-t_0}(t_0) }{1+n^z} \right]  \circledast \left[ \mathcal{A}h_{+}(t_0)+ \mathcal{B} h_{\times}(t_0)\right] \\ \nn
&+& \left[ \frac{ \Pi_{0,(m^z-1)(t_3-t_2)}(t_2) }{1-m^z}   +   \frac{ \Pi_{(m^z-1)(t_3-t_2),t_4-t_2}(t_2) }{1+m^z}\right]   \circledast  \left[ \mathcal{C}h_{+}(t_2)+ \mathcal{D} h_{\times}(t_2)\right]  \\ \nn
\eea

The ACABA travel time reads, starting from $t=t_0$, and using the new timestamps at each spacecraft $t_1'$, $t_2'$ and $t_4'$, 
\bea
t_4' - t_0 = &4L&       \\ \nn
&+& \left[   \frac{ \Pi_{0,(m^z-1)(t_1'-t_0)}(t_0) }{1-m^z}.  + \frac{ \Pi_{(m^z-1)(t_1'-t_0),t_2'-t_0}(t_0) }{1+m^z} \right]  \circledast \left[ \mathcal{C}h_{+}(t_0)+ \mathcal{D} h_{\times}(t_0)\right] \\ \nn
&+& \left[ \frac{ \Pi_{0,(n^z-1)(t_3'-t_2')}(t_2') }{1-n^z}   +   \frac{ \Pi_{(n^z-1)(t_3'-t_2'),t_4'-t_2'}(t_2') }{1+n^z}\right]   \circledast  \left[ \mathcal{A}h_{+}(t_2')+ \mathcal{B} h_{\times}(t_2')\right]  \\ \nn
\eea

The OPD (in terms of coordinate time) is then simply the difference of the two last formulas. The values depends on $t_0$ and writes $h(t_0)=t_4'-t_4$. It gives a very ugly result, but we can somewhat simplify it to allow for practical numerical application
with few additional assumptions. We will assume in the differential equation that $t_1\approx t_1'\approx L+t_0$, $t_2\approx t_2'\approx 2L+t_0$, $t_3\approx t_3'\approx 3L+t_0$,  and $t_4\approx t_4'\approx 4L+t_0$.. This assumption is realistic because the extra duration of the photon in an arm length caused by the GW is small with respect to the total duration $L$.
\bea
h(t_0) = && \left[   \frac{ \Pi_{0,(m^z-1)L}(t_0) }{1-m^z}.  + \frac{ \Pi_{(m^z-1)L,2L}(t_0) }{1+m^z} \right]  \circledast \left[ \mathcal{C}h_{+}(t_0)+ \mathcal{D} h_{\times}(t_0)\right]  \nn \\ \nn
&+& \left[ \frac{ \Pi_{0,(n^z-1)L}(t_2) }{1-n^z}   +   \frac{ \Pi_{(n^z-1)L,2L}(t_2) }{1+n^z}\right]   \circledast  \left[ \mathcal{A}h_{+}(t_2)+ \mathcal{B} h_{\times}(t_2)\right]  \\ \nn
&-& \left[   \frac{ \Pi_{0,(n^z-1)L}(t_0) }{1-n^z}.  + \frac{ \Pi_{(n^z-1)L,2L}(t_0) }{1+n^z} \right]  \circledast \left[ \mathcal{A}h_{+}(t_0)+ \mathcal{B} h_{\times}(t_0)\right] \\ \nn
&-& \left[ \frac{ \Pi_{0,(m^z-1)L}(t_2) }{1-m^z}   +   \frac{ \Pi_{(m^z-1)L,2L}(t_2) }{1+m^z}\right]   \circledast  \left[ \mathcal{C}h_{+}(t_2)+ \mathcal{D} h_{\times}(t_2)\right]  \\ 
\eea
It can then further simplify by using the Dirac operator $\delta(t-x)$, which, through convolution, amount to a time delay $x$:
\be
\begin{split}
h(t) =  & [ \delta(t)- \delta(t-2L) ]  \circledast   \left[   \frac{ \Pi_{0,(m^z-1)L}(t) }{1-m^z}.  + \frac{ \Pi_{(m^z-1)L,2L}(t) }{1+m^z} \right]   \circledast \left[ \mathcal{C}h_{+}(t)+ \mathcal{D} h_{\times}(t)\right]  \\ 
&-  [ \delta(t)- \delta(t-2L) ]  \circledast \left[   \frac{ \Pi_{0,(n^z-1)L}(t) }{1-n^z}.  + \frac{ \Pi_{(n^z-1)L,2L}(t) }{1+n^z} \right]  \circledast \left[ \mathcal{A}h_{+}(t)+ \mathcal{B} h_{\times}(t)\right]   \\
\end{split}
\label{eq:ht}
\ee
where $t$ now correspond to $t_0$, the time at which the photon left spacecraft $A$.

\begin{figure}
\centering
\includegraphics[width=0.75\textwidth]{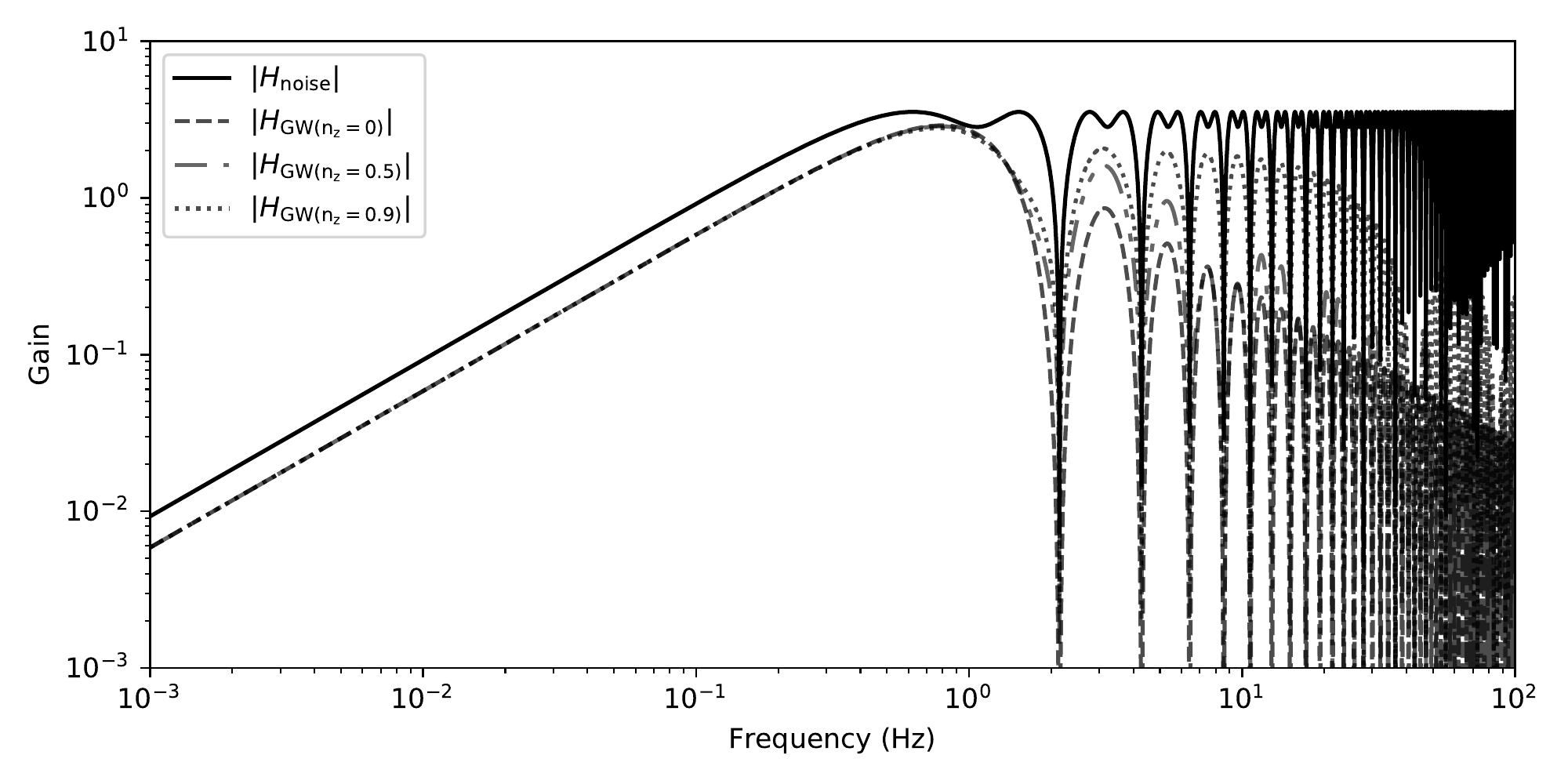}
\caption{ Amplitude of the transfer function of noise, $h_+$ and $h_\times$ components. The Sagnac measurement has a bandpass which frequency is centered on the lifetime of a photon inside an interferometric arm (4 times the distance between the satellites -- $\approx290\,000$\,km -- corresponding to 1\,Hz). The first cut-off frequency is due to the Sagnac configuration, and correspond to two arms length: $c/2L=2.1\,$Hz.
\label{fig:tf}}
\end{figure}

Then, using $\tilde{h}(f)$ as the Fourier transform of $h(t)$, we can write:
\begin{equation}
\begin{split}
\tilde{h}(f)/L=& H_{\rm GW (m^z)}(f) \times [
\mathcal{C}\tilde{h_+}(f)+\mathcal{D}\tilde{h_\times}(f)]\\
&- H_{\rm GW (n^z)}(f) \times [
\mathcal{A}\tilde{h_+}(f)+\mathcal{B}\tilde{h_\times}(f)]\\
\end{split}
\label{eq:tht_long}
\end{equation}
where $H_{\rm GW (n^z)}(f)$ is the transfer function for one arm of the detector:
\be
\begin{split}
H_{\rm GW (n^z)}(f)=&2\sin(2\pi L f) \times \\
&\left[ \mathrm{sinc}((n^z-1)Lf)\exp(-i \pi Lf) +\mathrm{sinc}((n^z+1)Lf)\exp( i \pi Lf) \right]\exp(-i\pi n^zLf)
\end{split}
\ee
where a constant phase term of value $L$ has been omitted. The first term, $2\sin(2\pi L f)$ corresponds to the Fourier transform of $ \delta(t)- \delta(t-2L)$. The second terms, $\mathrm{sinc}((n^z-1)Lf)\exp(-i \pi Lf) $ are the Fourier transfrom of the top-hat function ($\mathrm{sinc}(x)=\sin(\pi x)/(\pi x)$). The last term, $\exp(-i\pi n^zLf)$ correspond to a delay caused by the photon traveling along the $z$ axis, the propagation direction of the GW.

\begin{figure*}
\centering
\includegraphics[width=0.8\textwidth]{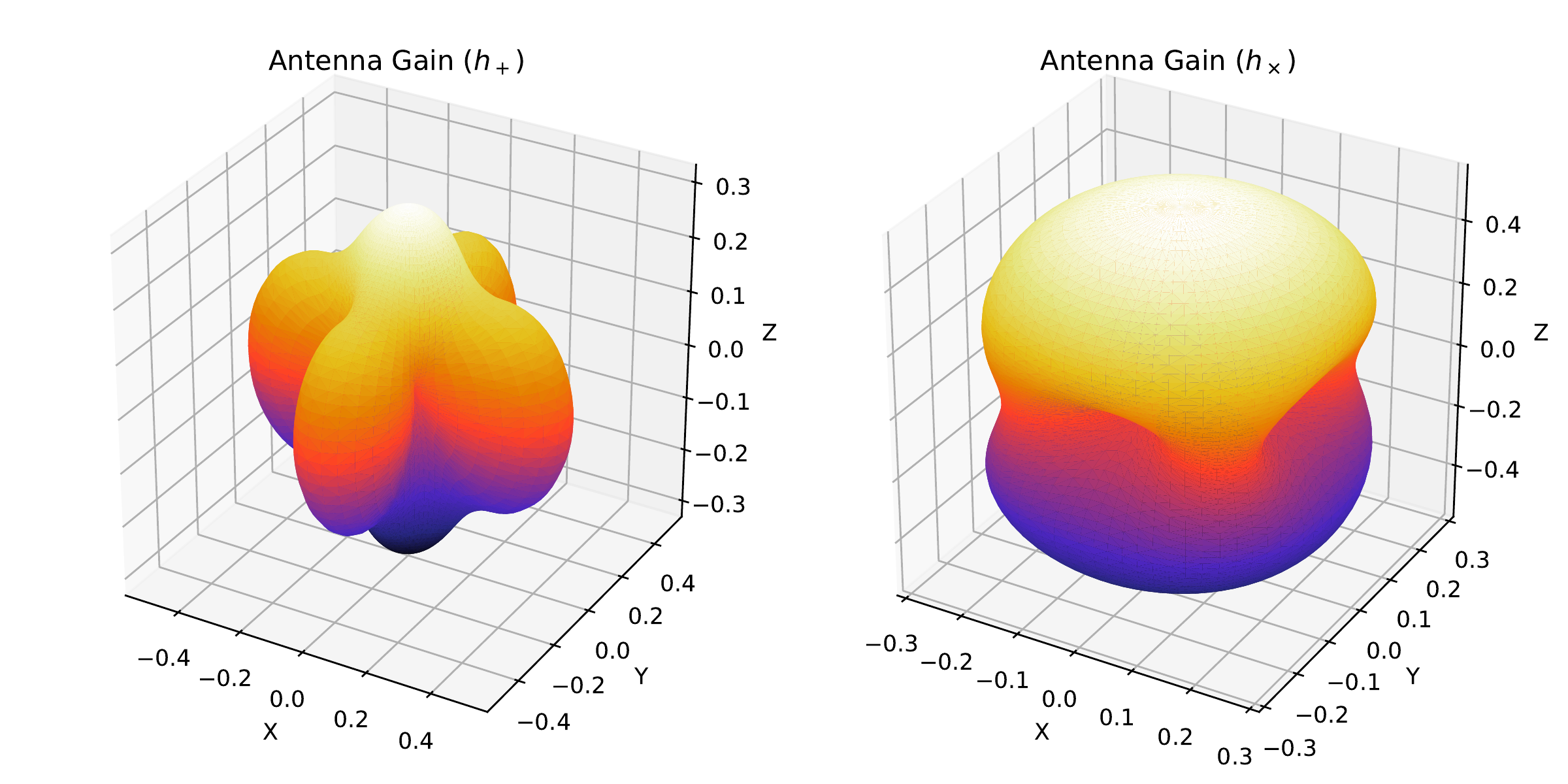}
\caption{ Interaction profile between the GW and the orbital configuation of the satellites. The GW is assumed to travel along the $z$ axis, with $x$ and $y$ corresponding to the axis for the $h_+$ polarization. The plots represent the orientation of the vector orthogonal to the plane of the equilateral configuration. The gains are averaged over all positions of the spacecrafts inside that plane. The amplitude are calculated from $\mathcal{C}-\mathcal{A}$ and $\mathcal{D}-\mathcal{B}$ (respectively $h_+$ and $h_\times$ polarizations) values as defined in Eq~(\ref{eq:ABCD}). The mean value for all sky positions are  $< \mathcal{C}-\mathcal{A} >_{\rm sky} = 0.29$ and $< \mathcal{D}-\mathcal{B} >_{\rm sky} = 0.33$. 
\label{fig:antennaGain}}
\end{figure*}

The transfer functions for $n_z= 0$, 0.5 and 0.9 are plotted Fig.~\ref{fig:tf}. It can be seen that for the operational frequencies of SAGE, which are below 2\,Hz, we can approximate that
\be
H_{\rm GW (n^z)}(f)\approx H_{\rm GW (n^z=0)}(f)=4\sin(2\pi L f) \mathrm{sinc}(2Lf)\,,
\label{eq:Hgw}
\ee
meaning that, for low frequency gravitational waves, we can neglect the effect of light propagation along the $z$ axis.
In that approximation, Eq.~(\ref{eq:tht_long}) becomes:
\begin{equation}
\tilde{h}(f)/L= H_{\rm GW (0)}(f) \times [
(\mathcal{C}-\mathcal{A})\tilde{h_+}(f)+(\mathcal{D}-\mathcal{B})\tilde{h_\times}(f)]\\
\label{eq:tht}
\end{equation}
the equations are therefore simpler since the effect of the geometry of the spacecraft configuration is simplified to 2 parameters: $\mathcal{C}-\mathcal{A}$ and $\mathcal{D}-\mathcal{B}$. These two parameters, also called "antenna gains" are plotted in Fig~\ref{fig:antennaGain} as a function of the vector orthogonal to both vectors $\bf n$ and $\bf m$ (in other words, orthogonal to the plane of the interferometer). The antenna gains are variable as a function of the direction of the incoming wave, and as a function of the polarization orientation of the gravitational wave. The gains values range between 0 and 0.5 for $\mathcal{C}-\mathcal{A}$ and 0 to 0.55 for $\mathcal{D}-\mathcal{B}$. The two gains, averaged over the entire sky, are approximately 0.29 for polarization $h_+$ and 0.33 for polarization $h_\times$. These numbers are computed from the following expression~\cite{moore15}
\be
\langle \mathcal{C}-\mathcal{A} \rangle_\mathrm{sky}^2 = \int_0^{2 \pi} \frac{\dd \psi}{2 \pi} \int_0^{2 \pi} \frac{\dd \varphi}{2 \pi} \int_0^{\pi} \frac{\sin \theta \dd \theta}{2} \left[\mathcal{C}(\theta,\varphi,\psi)-\mathcal{A}(\theta,\varphi)\right]^2
\ee
and similarly for $\mathcal{D}-\mathcal{B}$.

\subsection{Noise transfer function}
\label{sec:noise_tf}

A summary of the different noise that hinder the metrology measurement is presented in Table~\ref{tb:noise}. The final noise on the TDI measurement is the sum of the noise along each one of the eight pulses. Each one of these pulses are measurements at different times, and the noise error also must include these delay offset. If we assume the X configuration (ABACA and ACABA interferometric arms), the eight-pulse measurement includes 4 sources of noises: $n_{CA}(t)$, $n_{AC}(t)$, $n_{BA}(t)$ and $n_{AB}(t)$. The lower case denote the time domain, while the upper case corresponds to the power spectrum in the frequency domain. Then, $n_{\rm h}(t)$, the resulting noise on the TDI $h$ measurement,  writes:
\begin{equation}
\begin{split}
n_{\rm h}(t)&=n_{CA}(t)+n_{AC}(t -L)+n_{CA}(t-2L)+n_{BA}(t-2L)+n_{AB}(t-3L)+n_{BA}(t-4L)\\
&\ \ -n_{BA}(t)-n_{AB}(t-L)-n_{BA}(t-2L)-n_{CA}(t-2L)-n_{AC}(t-3L)-n_{CA}(t-4L)\\
&=[n_{CA}(t)-n_{BA}(t)]\circledast (1-\delta(4L))+[n_{AC}(t)-n_{AB}(t)]\circledast (\delta(L)-\delta(3L))\\
\end{split}
\end{equation}
where $L$ is the delay due to the propagation of the photon between two spacecrafts.

The variance $N_{\rm h}(f)$ can be obtained from the power spectrum of $n_{\rm h}(t)$, and using the Fourier transform of the Dirac functions ($D_L(f)=exp(2 \pi i L f)$):
\begin{equation}
\begin{split}
N_{\rm h}(f)=[N_{CA}(f)-N_{BA}(f)]|1-D_L^4|^2+[N_{AC}(f)-N_{AB}(f)]|D_L-D_L^3|^2\\
\end{split}
\end{equation}
In the assumption that each one of the $N_{CA}(f)$, $N_{AC}(f)$, $N_{BA}(f)$ and $N_{AB}(f)$ power spectrum are uncorrelated, stationary, and of equal amplitude $N(f)$, then:
\begin{equation}
N_{\rm h}(f)=|H_{\rm noise}(f)|^2 N(f)
\end{equation}
with $|H_{\rm noise}(f)|^2$ a real function which writes:
\begin{equation}
\begin{split}
|H_{\rm noise}(f)|^2&=2 |1-\exp(8\pi i f L)|^2 + 2 |1-\exp(4\pi i f L)|^2\\
			&=2 |2\sin(4\pi  f L)|^2 + 2 |2\sin(2\pi  f L)|^2\,.
\end{split}
\label{eq:Hnoise}
\end{equation}
The amplitude of the transfer function $|H_{\rm noise}(f)|$ is plotted in Fig.~\ref{fig:tf}. Its maximum is around 1\,Hz, corresponding to the maximum sensitivity of the interferometer. However, it decreases at low frequency as much as the interferometers response to GW decreases. This make it possible to observe GWs at frequencies as low as 10m\,Hz.

\subsection{Signal to noise and sensitivity function}

\begin{figure*}
\centering
\includegraphics[width=0.8\textwidth]{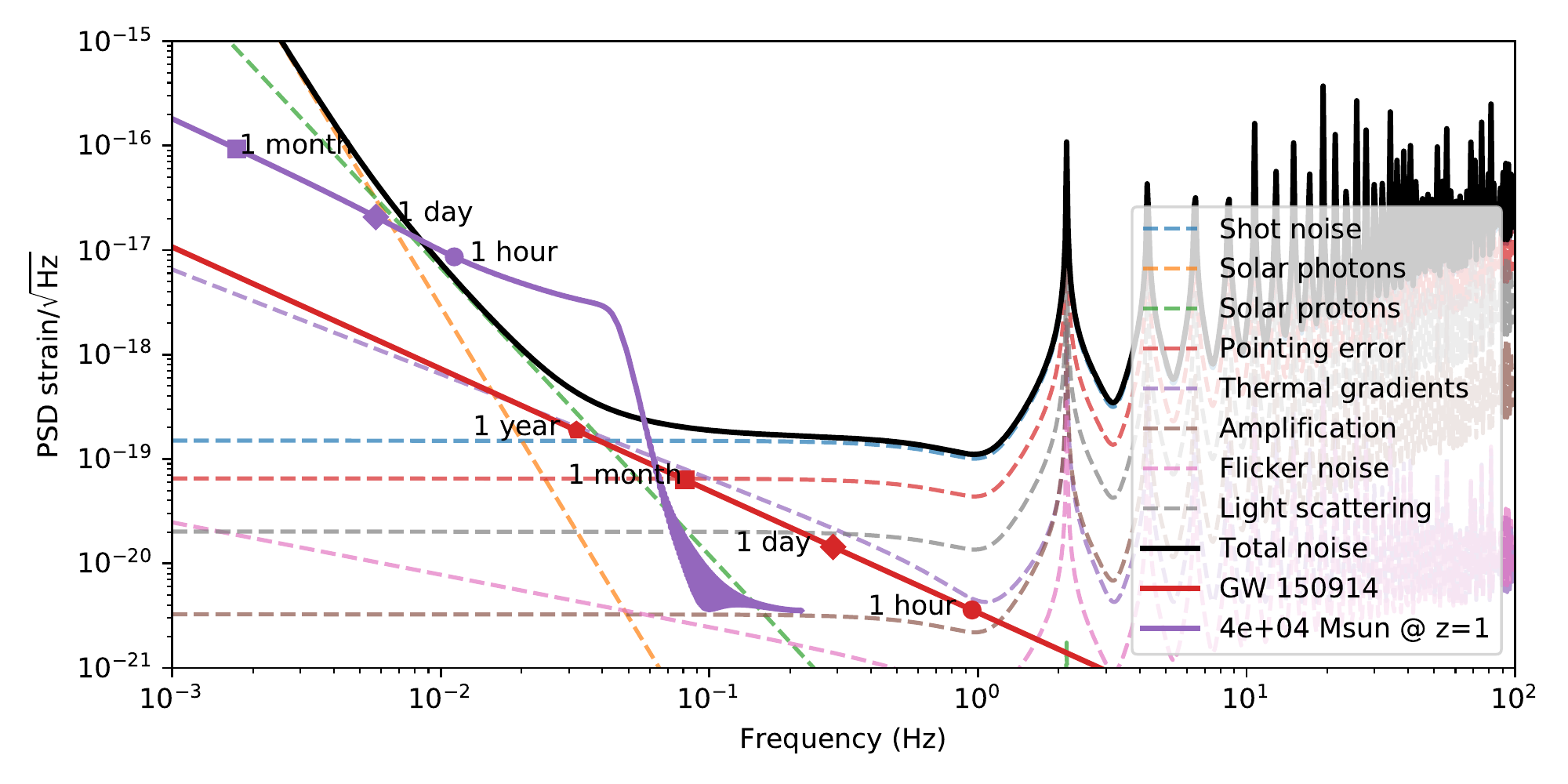}
\caption{Equivalent strain power density sensitivity of the SAGE interferometer for each source of noise presented in Section~\ref{sec:error}. The black line correspond to the total $1\sigma$ sensitivity of the interferometer. The 200\,mW laser and the small apertures makes it limited to a few $10^{-19}/\sqrt{\rm Hz}$ gravitational strain (blue dashed line). Over plotted are simulations of the gravitational strains caused by the mergers of a stellar mass and intermediate mass black holes. For the stellar mass black hole, we used the parameters of the first LIGO detection GW 150914 \cite{2016PhRvL.116x1102A}. The intermediate black holes merger correspond to two black holes of similar mass corresponding to a chirp mass of $4\times10^4\,M_\odot$.
\label{fig:strain_sens}}
\end{figure*}

The SNR is calculated from the average power spectrum over the range of frequencies accessible by the interferometer ($f_1=10$\,mHz and $f_1=2$\,Hz): 
\begin{equation}
SNR^2={3}\int_{f_1}^{f_2}\frac{4|\tilde{h}(f)|^2}{  N_{\rm h}(f)}\, df
\end{equation}
The factor 4 comes from the adopted convention of using one-sided power spectral densities. The factor 3 comes from the number of unequal-arm-length TDI measurements which can be used within an equilateral triangle. Hidden here is a factor 2/3 which applies both to the signal and noise (corresponding to the number of degrees of freedom with respect to the observable quantities).

Using the transfer function of the noise ($H_{\rm noise}(f)$) and of the interferometer ($H_{\rm GW}(f)$), we can expend the equation to introduce the independence to the noise power spectrum $N(f)$ and to the two gravitational polarizations $h_+$ and $h_\times$:
\begin{equation}
SNR^2=\int_{f_1}^{f_2}\frac{12|H_{\rm GW (0)}(f) |^2 L^2}{ |H_{\rm noise}(f)|^2 N(f)} \left|< \mathcal{C}-\mathcal{A} >_{\rm sky} \tilde{h_+}(f) + < \mathcal{D}-\mathcal{B} >_{\rm sky} \tilde{h_\times}(f) \right|^2 \, df\,,
\label{eq:snr2}
\end{equation}
where $N(f)$ is the sum of the power spectrum of all the noises enumerated in Table~\ref{tb:noise}. From this equation we can define the sensitivity function of the SAGE interferometer as:
\begin{equation}
S_{\rm h}(f)=\frac{ |H_{\rm noise}(f)|^2 N(f)}{12|H_{\rm GW (0)}(f) |^2 L^2}\,.
\label{eq:Sh}
\end{equation}

In Fig.~\ref{fig:strain_sens} are plotted the sensitivity $S_{\rm h}(f)$ as a function of the frequency for each noise. It can be seen that the system has a bandpass between 10\,mHz and 2\,Hz, as it was defined in the system parameter table~\ref{tb:param}. The sensitivity is maximum around 1\,Hz, corresponding to a gravitational strain of a few 10$^{-19}$. At low frequency, the performances are limited by the solar wind. At mid frequency, the performance are limited by the quantum noise of the laser. A high frequency, the system is limited by the arm length of the interferometer.

\subsection{Mass sensitivity}

Over plotted on Fig~\ref{fig:strain_sens} are two type of black-hole mergers simulated  using the PyCBC software \cite{Canton:2014ena}. The first one correspond to the GW\,150914 parameters: inclination of 140$^\circ$, two masses of 36 and 29$\,M_\odot$, at a distance of 410\,Mpc. The second one correspond to the merge of two identical solar black holes of total chirp mass $M_c=4\times10^4\,M_\odot$. To account for a redshift of $1$, the frequency is scaled by a factor 2, as well as the chirp mass. From this plot, it is clear that SAGE is lacking sensitivity to detect the inspiral of stellar black holes. However, sensitivity around $10^4$ or $10^5\,M_\odot$ is maximum.

The sensitivity limits (in optical distance or redshift) as a function of mass is plotted in Fig.~\ref{fig:massSensitivity}. The plot corresponds to the integration of the power spectrum of the signal to noise ratio between 10\,mHz and 2\,Hz, as established in Eq.~(\ref{eq:snr2}). The mergers were simulated using the PyCBC library assuming identical masses black hole mergers. The inclination were take to be 1\,rad equivalent to the average inclination value (assuming equipartition). Mass and frequencies are redshifted according to the distances.

The detection confidence is not trivially correlated to the SNR. However, it is accepted that an SNR above 8 is necessary to claim a detection. For mergers with masses between $10^4$ and $10^6\,M_\odot$, the iso-contour SNR has its mean at 800\,Mpc, equivalent to a redshift close to 0.12. The peak sensitivity is for chirp masses of $10^5\,M_\odot$ where IMBH of equal masses can be seen with an SNR of 8 at 1\,Gpc.

\section{Discussion}
\label{sec:summary}

Space projects results from difficult compromises between what is possible to do and what the scientists would like to observe. SAGE privilege the former on the later by looking at what can be done with the disruptive CubeSat technology \cite{NAP23503}. The main limitations comes from the quantum noise, the stability of the center of mass, and the solar wind. However, this sensitivity would enable the monitoring of $10\,^5$ solar mass black holes up to 1\,Gpc, and EMRI with a $10\,^4$ mass ratio up to 200\,Mpc (Section~\ref{sec:science}). This would open a new observation window, unprobed until the LISA mission.

The issue of the center of mass stability results in a several requirements. On top of it is thermal expansion, and the resulting need for being compact. This is why using the CubeSat format makes sense. But an additional advantage is the low cost access of containerized payload to space. For example, companies like spaceflight offer launch of 12U CubeSats to geostationary transfer orbit (GTO) for 2.75\,M\$. Access to GEO could also similarly be offered in the near future. 
Using a ratio of 10 between the satellite cost and the launch, we could estimate at 30\,M\$ the cost of the first satellite, including launch, which could be operated and tested prior to duplication. This cost cap could be maintained by the fact that the satellite would use only available, off-the-shelf, technologies. This would also allow a fast track development, which could also demonstrate technologies early on before the LISA mission.

Moreover, if CubeSats are relatively cheap to launch, they have also the advantage of being cheap to re-launch. It means that in case of success it can lead to an iterative approach where the satellites can be subsequently upgraded with new technologies. For example, quadrupling the laser power would enlarge the horizon by a factor 2 for stellar mass black holes. Also, a better control of the center of mass would increase the sensitivity in the intermediary mass range. Last, using a more sensitive sensor to calibrate the solar wind up to 99\% (instead of 90\%) would enable the observation of $10^6\,M_\odot$ mergers up to a few Gpc.
Such an iterative approach can be more adapted to discoveries with a new, and unknown, domain like mid frequency GW. 

Last, and maybe foremost, the complementary of multiple space based gravitational observatories is undeniable. This is already happening on the ground with the creation of several new observatories. First, the different location and orientation of the antennas allow a precise determinations of the polarization and location of the mergers. Second, in the case of observatories with different arm lengths, the detections at multiple frequencies allow complementary analysis of mergers at different stages of the inspirals.  Third, downtimes and the risk of failures inherent to space missions make it wise to have simultaneous projects. 

Black holes cosmic evolutions are mostly putative, as was demonstrated by the unpredicted high mass of the first observed stellar merger GW\,150914. The stochastic gravitational background of compact binary coalescence also keeps being refined on the observations of Virgo mergers \cite{2018PhRvL.120i1101A} and may well be very different at lower frequency. In the end, what is mostly likely to happen is that the models will be proved wrong. This can justify a different, but complementary, approach from LISA where a low cost mission has its place.

%\section*{ACKNOWLEDGMENTS}     
\ack
 SL acknowledges support from ERC starting grant No. 639248.
The VIRGO instrument onboard SoHO is a cooperative effort of scientists, engineers, and technicians, to whom we are indebted. SoHO is a project of international collaboration between ESA and NASA. Plot in Fig.~\ref{fig:inspi} was generated using the PyCBC software package
\cite{Canton:2014ena,Usman:2015kfa,Nitz:2017svb,alex_nitz_2018_1256897}.

\vspace{0.5cm}

\bibliographystyle{iopart-num}
\bibliography{sample}

\end{document}